\documentclass[prb,preprint,floatfix]{revtex4}

\usepackage{graphicx}
\usepackage{array}
\usepackage{amsmath,amsfonts,amssymb}
\usepackage[ansinew]{inputenc}
\usepackage{color}
\usepackage{calc}
\usepackage[normalem]{ulem}

\newcommand{\Figref}[1]{Fig.~\ref{#1}}

\begin{document}

\title{Submolecular-scale control of phototautomerization}

\author
{Anna Ros\l awska$^{1,2\ast}$, Katharina Kaiser$^1$, Michelangelo Romeo$^1$,  Elo\"ise Devaux$^3$, \\
 Fabrice Scheurer$^1$, St\'ephane Berciaud$^1$, Tom{\'a}\v{s} Neuman$^{4,5\ast}$, Guillaume Schull$^{1\ast}$\\
\normalsize{$^1$ Universit\'e de Strasbourg, CNRS, IPCMS, UMR 7504, F-67000 Strasbourg, France,} \\
\normalsize{$^2$ Max-Planck-Institut f{\"u}r Festk{\"o}rperforschung, D-70569, Stuttgart, Germany,} \\
\normalsize{$^3$ Universit\'e de Strasbourg, CNRS, ISIS, UMR 7006, F-67000, Strasbourg, France, }\\
\normalsize{$^4$ Institut des Sciences Mol\'eculaires d'Orsay (ISMO), UMR 8214, CNRS, Universit\'e Paris-Saclay, F-91405 Orsay Cedex, France.}\\
\normalsize{$^5$ Institute of Physics, Czech Academy of Sciences, Cukrovarnick\'{a} 10, 16200 Prague, Czech Republic.}\\
\altaffiliation{a.roslawska@fkf.mpg.de}
\altaffiliation{neuman@fzu.cz}\altaffiliation{schull@unistra.fr}}

\begin{abstract}
   
Many natural and artificial reactions including photosynthesis or photopolymerization are initiated by stimulating organic molecules into an excited state, which enables new reaction paths. Controlling light-matter interaction can influence this key concept of photochemistry, however, it remained a challenge to apply this strategy to control photochemical reactions at the atomic scale. Here, we profit from the extreme confinement of the electromagnetic field at the apex of a scanning tunneling microscope (STM) tip to drive and control the rate of a free-base phthalocyanine phototautomerization with submolecular precision. By tuning the laser excitation wavelength and choosing the STM tip position, we control the phototautomerization rate and the relative tautomer population. This sub-molecular optical control can be used to study any other photochemical processes.
       
\end{abstract}

\date{\today}

\maketitle

Photochemistry plays a central role in initiating and regulating fundamental natural and artificial reactions such as photosynthesis \cite{Mirkovic2017}, photochromism\cite{Bleger2015}, photopolymerization\cite{Corrigan2019}, transition-metal-complex photocatalysis\cite{Twilton2017} or phototautomerization\cite{Volker1976}. It relies on the optical stimulation of simple systems, mostly organic molecules, into an excited state where a new reaction path opens \cite{Mirkovic2017,Bleger2015,Corrigan2019,Volker1976,Twilton2017}.  
Controlling light-matter interactions and thereby influencing photochemistry has become a challenge for both fundamental and applied science.
Approaches relying on the manipulation of light have been proposed, such as the use of optical cavities that enhance the efficiency of chemical processes\cite{Hutchison2012}, as well as the use of metallic structures\cite{Mukherjee2013,Bockmann2016, Bockmann2018, Kazuma2018, Li2019} for the photo-generation of plasmons or hot electrons, that in turn induce chemical reactions.
So far, however, manipulating light at the atomic scale remains an unexplored photochemical strategy that may provide control over reactions with sub-molecular precision. 

Thanks to the extreme enhancement of the electromagnetic field at the apex of scanning tunnelling microscope (STM) tips, it has recently become possible to investigate fluorescence \cite{Qiu2003,Merino2015, Zhang2016, Imada2016, Doppagne2018, Kaiser2019, Doppagne2020, Yang2020, Roslawska2021, Cao2021, Imada2021, Imai-Imada2022, Roslawska2022, Dolezal2022, Imada2022}, Raman scattering \cite{Zhang2013, Lee2019, Liu2019, Li2021} or photocurrent \cite{Imai-Imada2022} with submolecular spatial resolution, \textit{i.e.}, far beyond what is attainable with other near-field and super-resolved far-field optical methods \cite{Orrit1990, Anger2006, Kuhn2006, Chen2017, Trebbia2022}.
Here, we build on this strategy and demonstrate that control over photoinduced intramolecular hydrogen transfer\cite{Volker1976, Imai-Imada2022} -- phototautomerization -- within a free-base phthalocyanine (H$_{2}$Pc) can be obtained by addressing a specific sub-part of the molecule with tip-enhanced laser irradiation. The phototautomerization rate and the relative tautomer population are controlled by tuning the laser excitation wavelength, and by choosing the STM tip position. 
The tautomerization mechanism is investigated on the basis of atomically resolved tip-enhanced photoluminescence (TEPL)\cite{Yang2020,Imada2021,Imai-Imada2022, Imada2022} spectroscopy and hyperspectral mapping, which are quantitatively supported by a comprehensive theoretical model, that unravel an excited-state mediated process. 
These results open the way to a new photochemical strategy where chemical reactions can be controlled with submolecular resolution.

\noindent  

\begin{figure}
  \includegraphics[width=\linewidth]{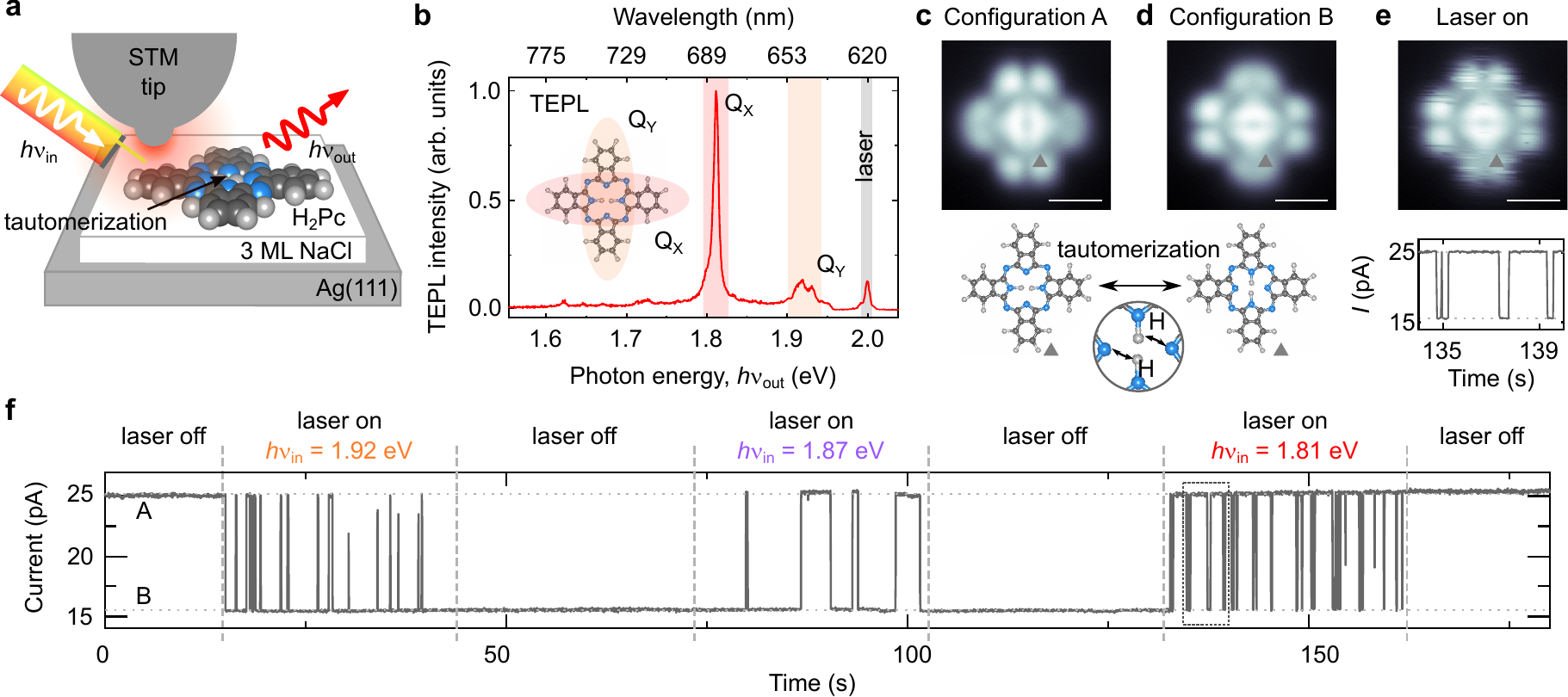}
  \caption{\label{fig1} \textbf{Phototautomerization of an individual molecule.} a) Scheme of the experiment. Incoming light of tunable energy $h\nu_{\rm in} $ excites an H$_{2}$Pc molecule leading to tautomerization and photoluminescence. The red shading indicates the extent of the plasmonic light enhancement provided by the picocavity. b) Tip-enhanced photoluminescence (TEPL) spectrum of H$_{2}$Pc, $h\nu_{\rm in}~=$~2.00~eV, integration time $t=$ 90 s, bias voltage $V=$ 0.55 V, tunnelling current $I=$ 2 pA, laser power $P\approx$~15~$\mu$W. Inset: sketch of the spatial arrangement of the Q$_{x}$ and Q$_{y}$ transition dipole moments. c-e) Constant-current STM images recorded for the two (configuration A and B) tautomers without (c,d) or with (e) laser illumination ($h\nu_{\rm in} $ = 1.81 eV, $P \approx$ 2 $\mu$W). Parameters: $V=$ 0.55 V, $I=$ 10 pA, scale bar: 1 nm. The chemical structure of both tautomers is presented in the bottom part of panels c) and d) together with a zoom into the molecular core where the hydrogen switching occurs. f) Current \textit{vs}. time trace recorded at constant height ($V=$ 0.55 V, $P \approx$ 2 $\mu$W) at the position marked by a grey triangle in e) in the alternating absence and presence of laser illumination. The incoming photons induce switching between two current values, which are assigned to configurations A and B, respectively, constituting a hallmark of tautomerization\cite{Liljeroth2007}. A section of the trace indicated by the dashed rectangle is presented in the bottom part of panel e).} 
\end{figure} 

Tautomerization is observed in many model molecules like free-base phthalocyanines\cite{Kugel2017a, Doppagne2020}, naphthalocyanines \cite{Liljeroth2007}, porphyrins \cite{Auwarter2012} and porphycenes\cite{Kumagai2013,Kumagai2013a, Bockmann2016,Bockmann2018}. In H$_{2}$Pc, this process consists in concerted or sequential\cite{Kugel2017a} switching of the two hydrogen atoms between two \textit{trans} configurations of the molecular centre
\cite{Doppagne2020}. This reaction can be induced by tunnelling current\cite{Liljeroth2007, Auwarter2012,Kumagai2013,Kugel2017a,Doppagne2020}, temperature\cite{Kumagai2013a} or photocarriers 
\cite{Bockmann2016,Bockmann2018} at a single molecule level and probed using STM \cite{Kugel2017a, Doppagne2020,Liljeroth2007,Auwarter2012,Kumagai2013,Kumagai2013a, Bockmann2016,Bockmann2018}.

A sketch of our experimental set-up is presented in \Figref{fig1}a. H$_{2}$Pc molecules are deposited on an Ag(111) single crystal partially covered with 3 monolayers (ML) of NaCl, a system that preserves the intrinsic electronic and optical properties of H$_{2}$Pc by preventing excited state quenching\cite{Zhang2016,Imada2016,Doppagne2018,Kaiser2019,Doppagne2020,Yang2020, Cao2021,Imada2021,Imai-Imada2022, Imada2022,Roslawska2022,Dolezal2022}. The experiments are done in a low-temperature (6 K) ultrahigh vacuum STM environment, which inhibits molecular diffusion. The molecules are optically excited by a tunable laser source focused on an STM Ag tip; the resulting photoluminescence signal is collected in the far field and guided to a detector located outside the vacuum chamber. More details on the experimental set-up can be found in Supplementary Information Section S1. 
\Figref{fig1}b shows a typical TEPL spectrum of H$_{2}$Pc excited with high-energy photons ($h\nu_{\rm in} =$ 2.00 eV). It shows two prominent features at $h\nu_{\rm out} =$ 1.81 eV and $h\nu_{\rm out} =$ 1.93 eV that can be assigned to the transitions from the two lowest energy excited states to the ground state, S$_{1}$ $\xrightarrow{}$ S$_{0}$  and S$_{2}$ $\xrightarrow{}$ S$_{0}$, usually referred to as Q$_{x}$ and Q$_{y}$, respectively \cite{Murray2011,Doppagne2020, Roslawska2022}. Spatially, the transition dipole moment of Q$_{x}$ is oriented along the two inner hydrogen atoms, and the one of Q$_{y}$ perpendicular, as shown in the inset of \Figref{fig1}b.
These two emission lines are accompanied by a series of low-energy vibronic peaks.

Thanks to its sensitivity to submolecular changes of the electronic structure, STM enables clear identification of the H$_{2}$Pc tautomers. \Figref{fig1}c,d are STM images of the lowest unoccupied molecular orbital (LUMO) density of H$_{2}$Pc that show two similar twofold symmetric patterns, rotated by 90$^{\circ}$, which can be assigned to the two different tautomers (bottom of \Figref{fig1}c,d). Upon laser illumination, the molecule is switching between the two configurations. This photoinduced effect can be visualized by locating the tip at a fixed position (marked in \Figref{fig1}e) and recording the tunnelling current trace (bottom of \Figref{fig1}e) under laser illumination ($h\nu_{\rm in} =$ 1.81 eV in this case). This trace exhibits two-level switching reflecting the different conductance of the two tautomer configurations for this tip position. Recording an STM image under illumination of the junction (\Figref{fig1}e) reveals now a fourfold symmetric pattern with noisy lines that are characteristic of the fast - as compared to the scanning speed - and continuous switching between the two \textit{trans} tautomers of H$_{2}$Pc. 
Here, we define as A (B) the configuration corresponding to the hydrogen atoms aligned horizontally (vertically) in the STM image. The two intrinsically identical configurations lead to different tunnelling current values for specific tip positions (\textit{e.g.}, markers in \Figref{fig1}c,d). This notation is consistent throughout this work.

To further visualize the influence of the illumination on the tautomerization, we maintain the tip at the same position and record a current trace for changing illumination energy (\Figref{fig1}f). At the beginning of the sequence ($t$ = 0), the laser is off and no switch is observed. This demonstrates that at a low bias ($V=$ 0.55 V), the tunnelling current does not drive tautomerization, rendering it a convenient non-perturbing probe of the photo-driven configuration of the system. Afterwards, we illuminate the junction with photons of different energies ($h\nu_{\rm in} =$ 1.92; 1.87; 1.81 eV) separated by periods during which the laser is switched off. One can see (\Figref{fig1}f) that the tautomerization rate depends on the excitation energy. Remarkably, we observe that the relative populations of the two tautomers also depend on the incoming photon energy: at high photon energy, configuration B is favoured, whereas at low energy it is configuration A.

\begin{figure}
  \includegraphics[width=0.38\linewidth]{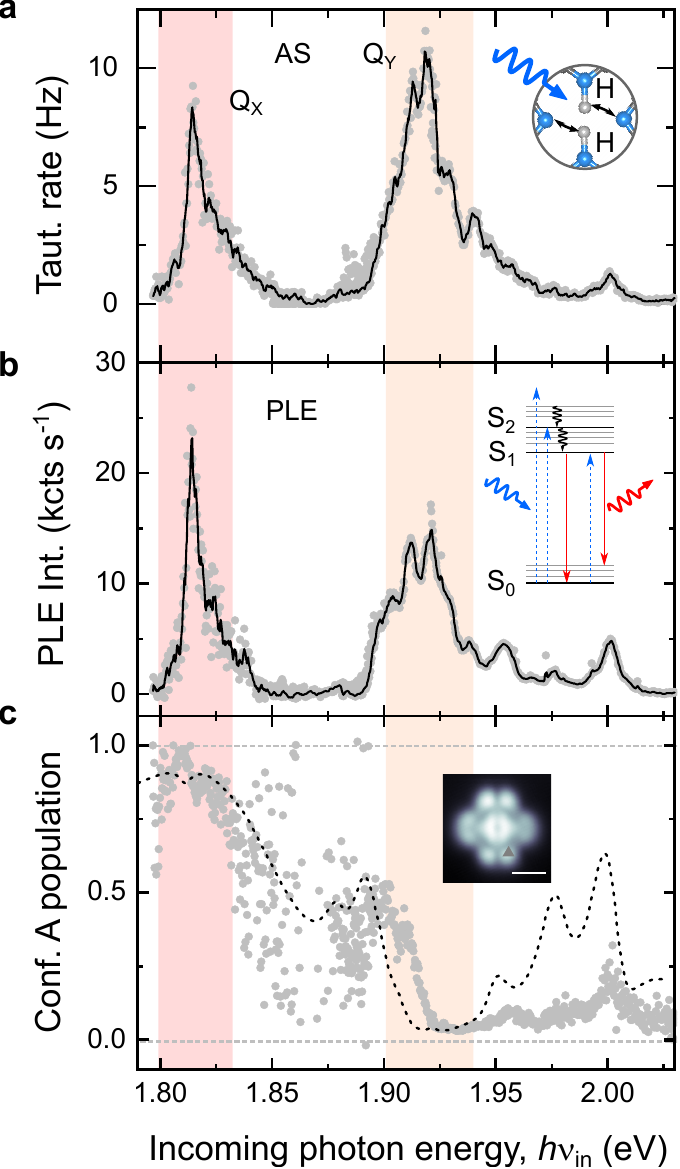}
  \caption{\label{fig2} \textbf{H$_{2}$Pc phototautomerization controlled by incoming photon energy.} a) Action spectroscopy (AS). The tautomerization rate as a function of the incoming photon energy is probed by monitoring the two-state switching in the the vertical displacement traces (closed feedback loop). b) Photoluminescence excitation (PLE) spectrum of H$_{2}$Pc. The inset shows the scheme of the PLE measurement: the molecule is excited \textit{via} a tunable laser, relaxes to the ground state of S$_1$ and eventually emits photons at the energy of the Q$_{x}$ transition whose intensity is monitored. Note that for $h\nu_{\rm in} <$ 1.85 eV we monitor the intensity of the vibrational emission. The measurements in a) and b) are recorded simultaneously. The acquisitions are performed in four energy intervals (1.80-1.86 eV; 1.85-1.90~eV; 1.88-1.93 eV; 1.92-2.03 eV) that allow correction for the chromatic aberration of the set-up (acquisition time per point, $t$ = 60 - 120 s, $P$ = 0.2 - 60 $\mu$W, $I$ = 5-10 pA, $V$ = 0.55 V). The shaded areas in a) and b) indicate the energy ranges corresponding to Q$_{x}$ and Q$_{y}$ transitions where both AS and PLE measurements show an increased signal. c) Configuration A population control by tuning the incoming photon energy. The data are calculated from the traces used to compute the tautomerization frequency presented in a). Inset: STM image (\Figref{fig1}c) of H$_{2}$Pc indicating the tip position (grey triangle) where the spectra have been recorded. The dashed line shows the configuration A population as calculated with our theoretical model.} 
\end{figure}

We first investigate the tautomerization rate (\Figref{fig2}a) as a continuous function of the incoming photon energy at a fixed tip position. 
Such a measurement, referred to as action spectroscopy (AS), is typically used to characterize photochemical processes\cite{Coohill1991}. We observe two switching rate maxima at around $h\nu_{\rm in} =$ 1.81 eV and $h\nu_{\rm in} =$ 1.92 eV (\Figref{fig2}a). We compare \Figref{fig2}a to a simultaneously recorded photoluminescence excitation (PLE) spectrum (\Figref{fig2}b), which consists in monitoring the photoluminescence intensity of Q$_x$ or vibronic peaks for different excitation wavelengths\cite{Orrit1990, Imada2021} (see Supplementary Information Section S1 for a detailed description).
Note that the PLE spectrum reflects the excitation spectrum of the chromophore, in contrast to a TEPL spectrum that yields information on the emission. In the PLE spectrum, we observe resonant features at the same energies as in the AS, corresponding to the transition energies of Q$_{x}$ (1.81 eV) and Q$_{y}$ (1.92 eV). This demonstrates that the tautomerization of H$_{2}$Pc on a thin insulator is an excited-state-mediated process\cite{Doppagne2020}, the generic reaction path in photochemical reactions, which can be driven resonantly by a tunable light source. This is in contrast with common STM-induced chemistry experiments where the path involves vibronic or charged states of a molecule excited by tunnelling or laser-generated hot electrons\cite{Kugel2017a,Li2019} whose energy cannot be well controlled.
We observe the same behaviour for a chemically different species, HPc$^-$ (see Supplementary Information Section S2). In addition, we find an increased tautomerization rate and photoluminescence signal for $h\nu_{\rm in} =$ 2.00 eV, which we attribute to an efficient coupling to specific vibrational modes of H$_{2}$Pc (see Supplementary Information Section S3 for more details).

Besides controlling the tautomerization rate, the incoming photon energy also modulates the relative population of configurations A and B (\Figref{fig2}c). The populations are determined from the vertical displacement traces (closed feedback loop) recorded under illumination (see Supplementary Information Section S1). At excitation energies resonant with Q$_{x}$ (1.80 eV $ <h\nu_{\rm in} <$ 1.83 eV) we find that the molecule spends the majority of the time in configuration A. This reflects a  higher probability of the B $\xrightarrow{}$ A transition to occur compared to the A $\xrightarrow{}$ B one, as can be deduced from \Figref{fig1}f. For 1.83 eV $< h\nu_{\rm in} <$ 1.92 eV we observe an intermediate regime, where configuration A is decreasingly favoured, a value that reaches minimum for excitation energies resonant with Q$_{y}$ ($h\nu_{\rm in} \approx$ 1.92 eV). Besides, we also observe additional features at excitation energies $h\nu_{\rm in} = $ 1.90 eV and $h\nu_{\rm in} > $ 1.95 eV that can be linked to the resonant excitation of vibrational modes of Q$_{x}$ resulting in an increase of the configuration A population. These trends are well reproduced by a rate equation model (dashed curve in \Figref{fig2}c), which takes into account the vibrational modes of H$_{2}$Pc by considering their Franck-Condon activity. For simplicity, we neglect possible effects of non-adiabatic coupling of the molecule's excited states that could be expected due to the small energy separation between the Q$_x$ and Q$_y$ transitions and could give rise to e.g. Herzberg-Teller activity of specific vibrational modes. The model is discussed later in more detail. We remark that the tautomerization rate and the relative configuration population are not directly linked, that is, a faster tautomerization rate does not necessarily lead to an increased population of a given configuration (\Figref{fig2}a,c).

\begin{figure}
  \includegraphics[width=\linewidth]{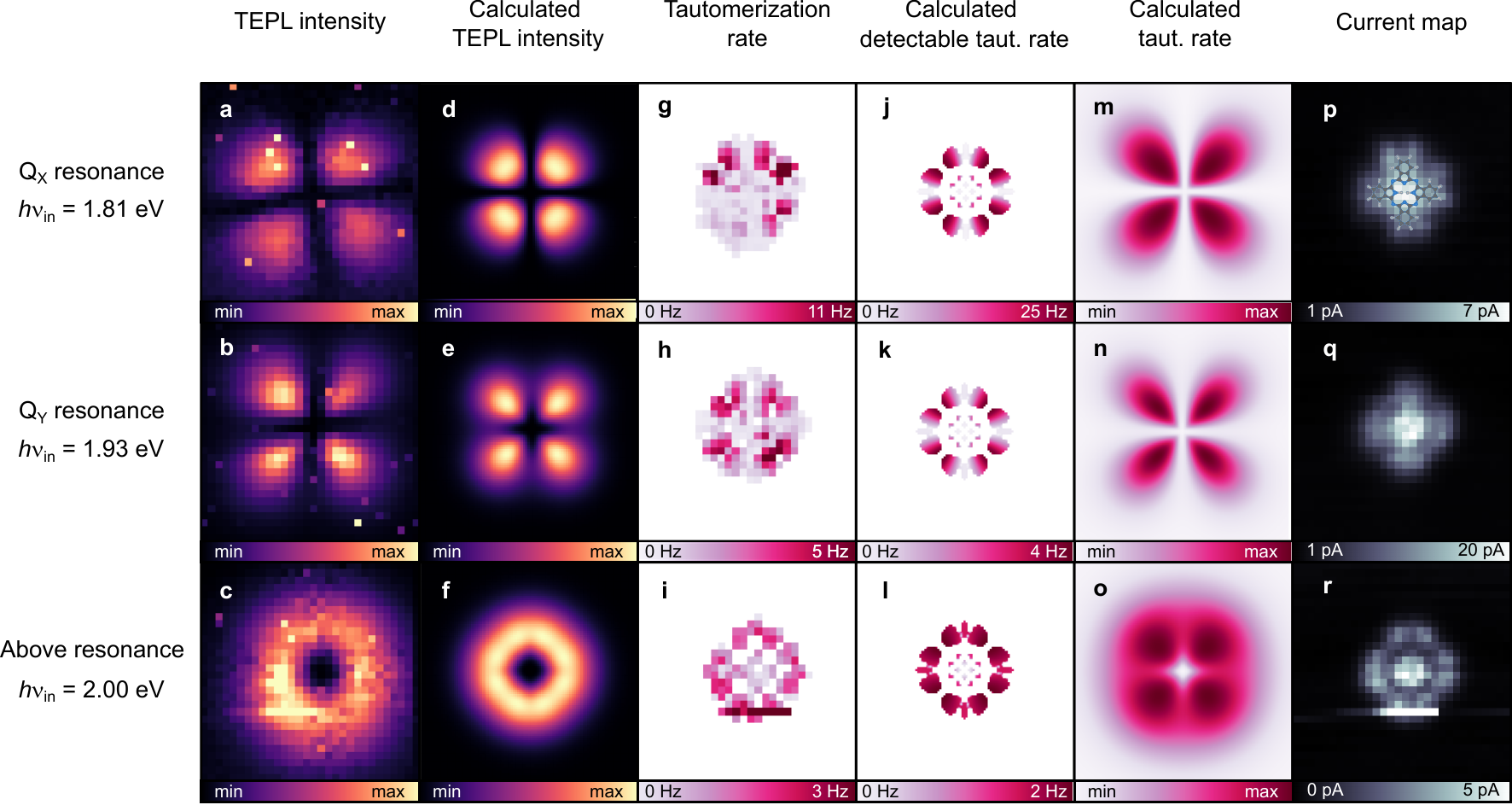}
  \caption{\label{fig3} \textbf{Submolecular control of the phototautomerization rate by local plasmonic fields.} a)-c) TEPL maps recorded with open feedback loop at different excitation energies ($h\nu_{\rm in}$). The intensities in the maps in a) (b, c) are integrated using the $h\nu_{\rm out}$ = 1.64-1.65 eV ($h\nu_{\rm out}$ = 1.80-1.82 eV) ranges, $P < 25~\mu$W. d)-f) Calculated TEPL maps. g)-i) Tautomerization rate as a function of the tip position. The patterns observed in the maps reflect a complex interplay between tautomerization and excitation/emission probability. j)-l) Calculated detectable tautomerization rate maps. 
  They are obtained from the maps in m)-o) by setting their value to zero when the tunnelling current passing through the molecule does not significantly differ between the two tautomer configurations. In g)-l) the white colour reflects points where no switch is detected.  m)-o) Calculated tautomerization rate maps. p)-r) STM constant height images recorded simultaneously with TEPL and tautomerization rate maps ($V$ = 0.55 V). The molecular structure (to scale) is overlaid in p). The size of all maps is $4.4 \times 4.4$ nm$^{2}$.} 
\end{figure}

Having characterized the phototautomerization at a given tip position, we explore now how we can control the tautomerization reaction with submolecular precision by tuning the coupling between the molecular exciton and the plasmons confined at the tip-sample picocavity\cite{Urbieta2018,Roslawska2022}. To do so, we simultaneously record photoluminescence and tautomerization rate maps for different excitation energies (\Figref{fig3}) in the constant height mode (bias voltage $V = 0.55$ V). Figures \ref{fig3}a-c show TEPL maps of H$_{2}$Pc for $h\nu_{\rm in}$ = 1.81, 1.93 and 2.00 eV, recorded by acquiring optical spectra at each pixel and integrating the intensity over the Q$_{x}$ transition (b,c) or its vibronic peaks (a). These experimental maps are complemented by simulations (\Figref{fig3}d-f). Simultaneously, we monitor the tunnelling current and extract the tautomerization rate, which is presented in \Figref{fig3}g-i and compared to simulations in \Figref{fig3}j-o. The lateral extension of these maps is limited to the region where the tunnelling current (\Figref{fig3}p-r) can be detected. In Supplementary Information, we provide more technical details on the measurement (Section S1) and additional TEPL maps (Section S4).

Under excitation resonant with either Q$_{x}$ ($h\nu_{\rm in}$ = 1.81 eV, \Figref{fig3}a) or Q$_{y}$ ($h\nu_{\rm in}$ = 1.93 eV, \Figref{fig3}b), the TEPL maps exhibit four bright lobes separated by a dark cross aligned with the molecular backbone (overlaid in \Figref{fig3}p). In contrast, the spatial distribution of the photon intensity for the non-resonant excitation ($h\nu_{\rm in}$ = 2.00 eV, \Figref{fig3}c) takes a doughnut-like shape. The simultaneously recorded tautomerization rate maps show similar patterns. At resonance (\Figref{fig3}g,h), the tautomerization rate exhibits a symmetric four-lobe shape and is lowest along the axes defined by the molecular backbone. Above resonance (\Figref{fig3}i), it exhibits a more radially homogenous doughnut-like shape. 
Eventually, comparing the TEPL, tautomerization and current spatial distributions, we observe that the incoming photons may excite the molecule at lateral distances for which electrons can no longer tunnel into the molecule (further discussion can be found in Supplementary Information Section S5).

The patterns observed in TEPL and tautomerization rate maps
can be explained by an interplay between the optical properties of the molecule and the tautomerization process. At $h\nu_{\rm in}$ = 1.81 eV, the excitation of Q$_{x}$ is strongly favoured, except when the tip is located on the perpendicular bisector of the Q$_{x}$ dipole. In this case, the coupling with the picovavity plasmon is nearly zero leading to the reduction of the TEPL signal\cite{Yang2020} (\Figref{fig3}a). In contrast, TEPL should be strongly enhanced when the tip is located on top of the extremity of the Q$_{x}$ dipole. This configuration, however, favours a fast tautomerization reaction, eventually locking the molecule in a configuration where the tip is again located at the perpendicular bisector of the 90$^{\circ}$-switched Q$_{x}$ dipole (as in \Figref{fig1}c). The locking in a specific configuration is also observed in \Figref{fig2}c for excitation energies resonant with Q$_{x}$ when configuration A is favoured.  Due to this mechanism, both the TEPL and tautomerization rate maps feature a dark cross aligned with the molecular axes (\Figref{fig3}a,g). Hence, the TEPL and the tautomerization rate are intense only where excitation and emission of the Q$_{x}$ dipoles (of both tautomers) are allowed. This condition results in the 4-lobe pattern with the maxima at 45$^{\circ}$ from the dipole axis. At $h\nu_{\rm in}$ = 1.93 eV (\Figref{fig3}b,h), we follow a very similar reasoning. Now, the excitation of Q$_{y}$ is possible, except when the tip is located on the perpendicular bisector of the Q$_{y}$ dipole. This results again in locking the molecule in a given configuration and leads to a very similar pattern as for $h\nu_{\rm in}$ = 1.81 eV. For $h\nu_{\rm in}$ = 2.00 eV, angularly homogeneous TEPL and tautomerization rate maps are observed (\Figref{fig3}c,i) revealing approximately equivalent excitation probabilities of Q$_{x}$ and Q$_{y}$ dipoles. We propose (see details in Supplementary Information Section S3) that this effect reflects the coupling of the incoming light to intense vibrational modes of H$_2$Pc. Further discussion on the features observed in the maps in \Figref{fig3} can be found in Supplementary Information Section S6.

To support this interpretation, we performed an extensive theoretical analysis based on a model (see Supplementary Information Sections S7-S9 for details) accounting for the effect of (i) tip-position-dependent plasmon-enhanced photoexcitation of the molecule, (ii) plasmon-enhanced spontaneous emission, (iii) vibronic structure of the molecular excitations derived from the vibrational Franck-Condon activity, and (iv) tautomerization dynamics driven by an intermediate electronic state of the molecule.  We solve the rate equations and generate the tip-position-dependent TEPL intensity  
(Fig.\,\ref{fig3}d-f) and tautomerization rates (Fig.\,\ref{fig3}j-o). The calculated absolute tautomerization rate maps (Fig.\,\ref{fig3}m-o) demonstrate that hydrogen switching occurs even when the tip is located relatively far from the molecule. To account for the spatially restricted sensitivity of the experiment, we plot the data originally shown in \Figref{fig3}m-o as non-zero only in regions where the magnitude of the difference in the density of states $\left ||\psi^{\rm A}_{\rm LUMO}|^2-|\psi^{\rm B}_{\rm LUMO}|^2\right|$ exceeds a predefined threshold value (\Figref{fig3}j-l, see details in Supplementary Information Section S9).
The resulting theoretical maps are in excellent agreement with the experiment. 

For the resonant excitation of Q$_x$ (\Figref{fig3}d), the model yields lobes that are distinctly less elongated in the radial direction than for the resonant Q$_y$ (\Figref{fig3}e) excitation, a behaviour one can also identify in the experimental maps. This specific characteristic suggests a triplet-mediated phototautomerization reaction path. Indeed, other paths, involving for example vibrational levels of the ground state, would lead to similarly elongated lobes in the  Q$_x$ (\Figref{fig3}a) and Q$_y$ (\Figref{fig3}b) TEPL maps (see Supplementary Information Sections S7 and S10). Overall, this theoretical analysis shows that submolecular photoluminescence mapping at different excitation energies is a powerful tool to study light-induced reaction mechanisms in a single molecule.

\begin{figure}
 \includegraphics{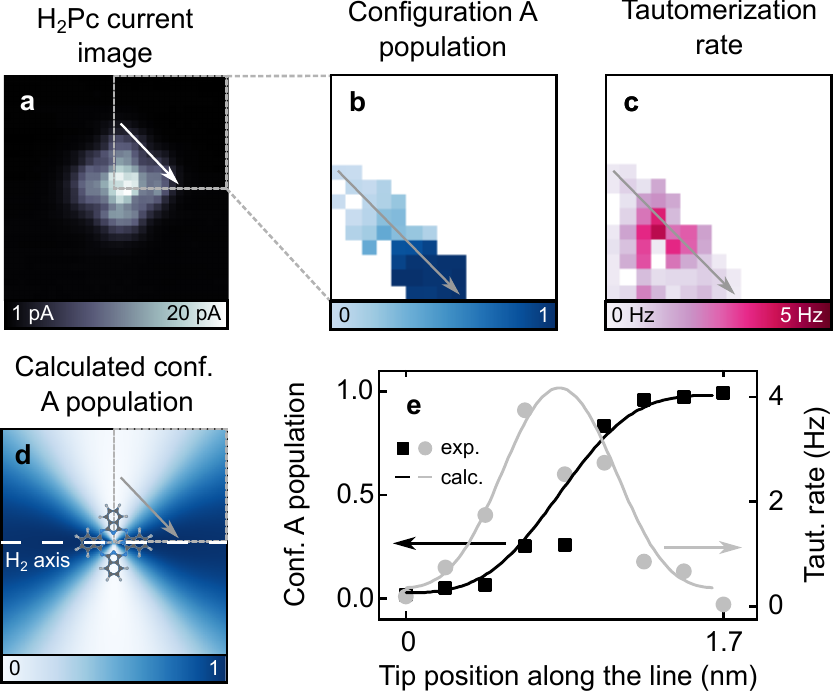}
 \caption{\label{fig4} \textbf{Submolecular control of the tautomer configuration population.} a) STM constant height image, $V$ = 0.55 V. b) Close-up view (marked by a dashed square in a) ) of the configuration A population and c) of the tautomerization rate. d) Calculated configuration A population as a function of tip position. The molecular backbone (to scale) is overlaid. e) Line-cuts of data in b-d taken along the arrows indicated in a-d. The solid lines are calculated using the rate-equation model presented in this work. The grey line is the calculated detectable tautomerization rate. The slight mismatch between the experimental and theoretical grey curves in (e) is related to  small inaccuracies in estimating the experimental tautomerization rate when the current of A and B configurations are close (see Supplementary Information Section S1 for details). Illumination parameters, $h\nu_{\rm in} =$ 1.93 eV, $P =$ 18  $\mu$W. Size of maps in a,d): $4.4 \times 4.4$ nm$^{2}$.} 
\end{figure}   

Finally, in \Figref{fig4} we show that the relative population of configurations A and B can be controlled with submolecular precision (dashed square in \Figref{fig4}a). In \Figref{fig4}b, we present the tautomer configuration A population and in \Figref{fig4}c the tautomerization rate as a function of the tip position, which are recorded under illumination at $h\nu_{\rm in} =$ 1.93 eV (see Supplementary Information Section S1). \Figref{fig4}b reveals that configuration A is favoured (disfavoured) when the tip is located on top of the right (top) benzopyrrole of the molecule, a behaviour that is well reproduced by our theoretical model (\Figref{fig4}d). 
Remarkably, the configuration population follows a different pattern than the tautomerization rate, as emphasised in the line cut (\Figref{fig4}e), highlighting the possibility of versatile tuning of the reaction rate and its outcome.

Overall, we have demonstrated that an excited-state driven photochemical process can be induced with atomic-scale control in a model molecular system. This is possible thanks to the intense field confinement occurring at the apex of a plasmonic STM tip, which is used to address a specific transition dipole within a molecule. We show control over the tautomerization rate and the tautomer populations by controllably addressing different submolecular units and tuning the excitation energy. The reversibility of this reaction, together with the extreme sensitivity of the STM to minute changes in the structure of a molecule, allowed us to follow the photoreaction in detail and explore different parameters driving the process. Our approach can readily be applied to induce and control other reversible reactions such as photochromism\cite{Comstock2007, Bazarnik2011, Bleger2015, Nacci2018} or on-surface photopolymerization\cite{Miura2003,Para2018,Clair2019,Grossmann2021} with submolecular precision, and may be extended to the ultrafast time domain\cite{Garg2019, Peller2020} to access short-lived intermediate states. While the use of a reversible model system was necessary to validate the approach, our concept is general and applies to any photochemical reaction, regardless of its reversible character. 
In future works, the sub-nanometric control reported here will be beneficial for site-selective photochemistry and may provide access to previously unavailable chemical compounds\cite{Clair2019}.

\noindent

\section*{Acknowledgements}
We would like to thank Virginie Speisser and Halit Sumar for technical support, Guillaume Rogez for help with UV-vis spectroscopy, Alex Boeglin, Luis E. Parra L\'opez, Song Jiang, \'Oscar Jover Arrate and Andrey Borissov for discussions. This project has received funding from the European Research Council (ERC) under the European Union's Horizon 2020 research and innovation program (grant agreement No 771850), the European Union's Horizon 2020 research and innovation program under the Marie Sk\l{o}dowska-Curie grant agreement No 894434, and the SNSF under the Postdoc.Mobility grant agreement No 206912. This work is supported by ``Investissements d'Avenir'' LabEx PALM (ANR-10-LABX-0039-PALM). TN acknowledges the Lumina Quaeruntur fellowship of the Czech Academy of Sciences. Computational resources were supplied by the project "e-Infrastruktura CZ" (e-INFRA CZ LM2018140) supported by the Ministry of Education, Youth and Sports of the Czech Republic. We acknowledge financial support from the Agence Nationale de la Recherche under grant ATOEMS ANR-20-CE24-0010.

\bibliographystyle{naturemag}

\bibliography{tauto_ref}

\end{document}

\title{Supplementary information for\\
Submolecular-scale control of phototautomerization}

\author
{Anna Ros\l awska$^{1,2\ast}$, Katharina Kaiser$^1$, Michelangelo Romeo$^1$,  Elo\"ise Devaux$^3$, \\
 Fabrice Scheurer$^1$, St\'ephane Berciaud$^1$, Tom{\'a}\v{s} Neuman$^{4,5\ast}$, Guillaume Schull$^{1\ast}$\\
\normalsize{$^1$ Universit\'e de Strasbourg, CNRS, IPCMS, UMR 7504, F-67000 Strasbourg, France,} \\
\normalsize{$^2$ Max-Planck-Institut f{\"u}r Festk{\"o}rperforschung, D-70569, Stuttgart, Germany,} \\
\normalsize{$^3$ Universit\'e de Strasbourg, CNRS, ISIS, UMR 7006, F-67000, Strasbourg, France, }\\
\normalsize{$^4$ Institut des Sciences Mol\'eculaires d'Orsay (ISMO), UMR 8214, CNRS, Universit\'e Paris-Saclay, F-91405 Orsay Cedex, France.}\\
\normalsize{$^5$ Institute of Physics, Czech Academy of Sciences, Cukrovarnick\'{a} 10, 16200 Prague, Czech Republic.}\\
\altaffiliation{a.roslawska@fkf.mpg.de}
\altaffiliation{neuman@fzu.cz}\altaffiliation{schull@unistra.fr}}

\maketitle

\tableofcontents

\newpage
\noindent

\subsection{S1 -- Experimental methods}

\textbf{Sample and tip preparation}

\begin{figure*}[!th]
  \includegraphics{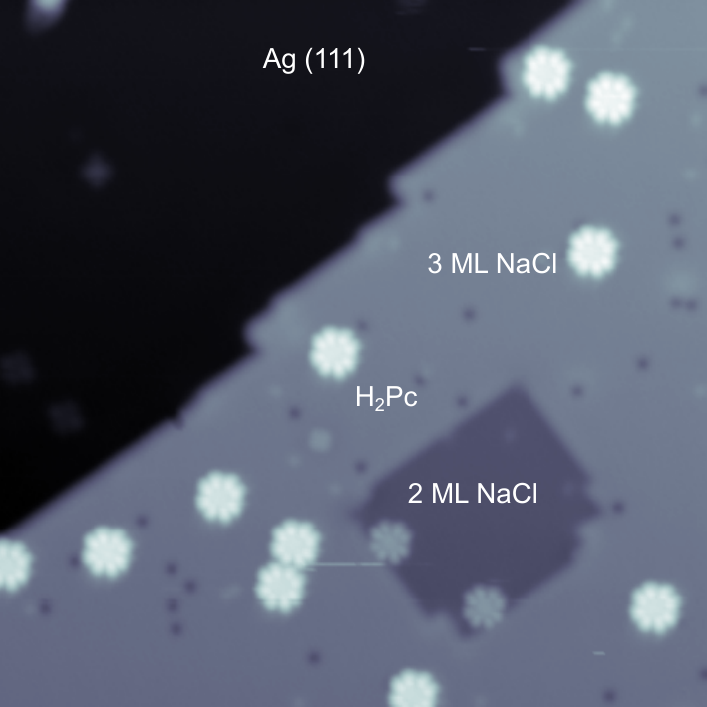}
 \caption{\label{SI_scan} STM overview image of the H$_{2}$Pc/NaCl/Ag(111) sample. Size: 36.5 $\times$ 36.5 nm$^{2}$, $V$ = -2.5 V, $I$ = 5 pA. } 
\end{figure*}
 
The Ag(111) substrate is prepared by successive argon-ion sputtering and annealing cycles. Approximately 0.5 monolayer NaCl is thermally sublimed onto Ag(111) kept at room temperature. Square bilayers and trilayers are formed after a mild post-annealing step. Free-base phthalocyanine (H$_{2}$Pc, Sigma-Aldrich) molecules are sublimed on the cold (6 K) sample located in the scanning tunneling microscope (STM) head. An image of the surface is presented in \Figref{SI_scan}. We use electrochemically etched Ag tips \cite{Zhang2011, Ramanauskaite2018}, which are either Ar-ion sputtered and annealed in ultra-high vacuum (UHV) or treated with focused ion beam (FIB) etching. The plasmonic response of the tips is optimized by voltage pulses and gentle indentations into the sample to reach a strong plasmonic resonance in the range of 1.8-2.0 eV as probed by STM-induced luminescence (STML).

To perform FIB etching we use an Auriga cross-beam system (Zeiss), allowing switching from ion beam milling to electron beam imaging to control the etching process. 
We use successive annular milling steps with currents varying from 20 nA (for rough etching) to 20 pA (for polishing), with an accelerating voltage of 30 kV. The inner and outer radii are chosen after an inspection under SEM of the initial dimensions of the tip. This sequence, lasting a couple of hours, results in a tip that is smooth over nearly 20 $\mu$m length, with an apex diameter of 50 nm.

\vspace{10pt}
\textbf{STM and optical set-up} 

\begin{figure*}[!h]
  \includegraphics[width=0.8\linewidth]{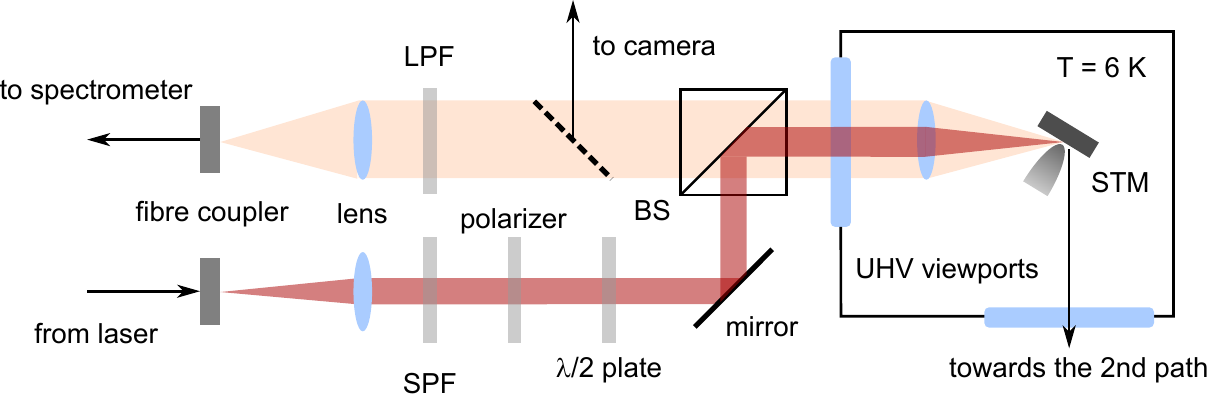}
  
 \caption{\label{SI_setup} Optical set-up used in the experiment. SPF - shortpass filter, LPF - longpass filter, BS - beamsplitter.} 
\end{figure*}

The data are acquired with a low-temperature (6 K), UHV Unisoku USM1400 set-up with optical access provided by two adjustable lenses (numerical aperture, NA = 0.55 each) located in the cold STM head (\Figref{SI_setup}). A tunable laser source, composed of a supercontinuum laser (NKT Photonics SuperK Extreme EXW-6) combined with an acousto-optic tunable filter (AOTF, NKT Photonics SuperK SELECT) provides narrow-band light (FWHM 0.5-1.85 nm) in the 430 - 695 nm range. Note that the laser linewidth may contribute to the additional broadening of the measured spectral features. The seed laser pulse duration is 5 ps and the repetition rate is 78 MHz. Note that despite using a pulsed laser, we treat it as a tunable continuous light source, as justified by the power dependence (see below), which does not show non-linear effects. The laser output is directed towards a viewport of the STM (\Figref{SI_setup}) and then focused onto the tip apex. 

The light emitted or scattered from the tip-sample junction is collected, filtered and directed towards a single-pass monochromator coupled to a liquid nitrogen-cooled CCD array (Princeton Instruments). 
Overall, we estimate the detection efficiency of the set-up to be 10 \%. The detection is performed in two configurations, either using the same output as the one used for the excitation or the one of the second lens located in the STM head. In the latter configuration, the optical detection path is placed on a different optical table. A sketch of the set-up configuration comprising one optical table is provided in \Figref{SI_setup}. 
In addition, a switchable mirror is implemented in the detection path so that the preliminary optical alignment can be performed with a camera. The final alignment is done directly on the tip-enhanced photoluminescence (TEPL) signal (in tunnelling contact) by carefully manipulating the positions of the mirrors and lenses located in the ambient until the maximum of the detected signal is reached. 
The TEPL maps presented in this work are recorded in the constant height mode (the feedback loop is disabled). The residual drift of the STM is corrected by employing the "atom tracking" feature of the Nanonis SPM controller several times during the measurement (in between spectra acquisitions).

\vspace{10pt}

\textbf{Power dependence}

\begin{figure*}[!h]
  \includegraphics[width=\linewidth]{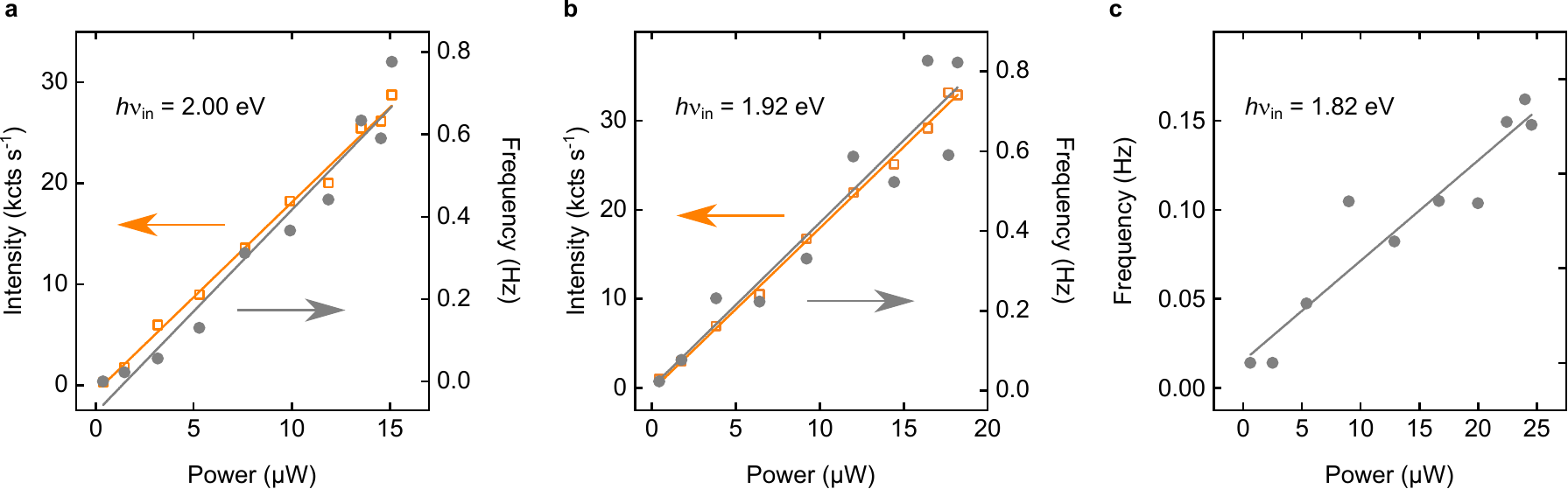}
  
 \caption{\label{SI_power} TEPL intensity (orange rectangles) and tautomerization frequency (grey circles) of H$_{2}$Pc recorded as a function of the laser power for three different excitation energies. All measurements are recorded at the extremity of a pyrrole unit. Integration range of the TEPL spectra: 1.79-1-82 eV, $V =$ 0.55 V, current set-point $I$ = 2 pA, the light intensity is corrected for the detection efficiency. The solid lines are linear fits to the data.} 
\end{figure*}

The optical power is regulated directly by the laser controller and measured in front of the UHV viewport. Note that the power value measured here provides only the upper bound for the power eventually delivered to the tunnel junction, as this value (and thus absorption events) depends on the final optical alignment in the low-temperature UHV STM head. Besides, the absorption (and emission) probability also depends strongly on the plasmonic enhancement provided by the tip. This may lead, for instance, to observing similar tautomerization rates and TEPL intensity for measured power values spanning $1 ~\mu \rm{W}< P < 50~\mu$W range. 
Note that, in our case, the maximum power that can be applied is eventually determined by the H$_2$Pc deprotonation (see Section S2), which is usually occurring at excitation intensities corresponding to tautomerization rates beyond 10 Hz.

Further, we studied how the tautomerization rate and TEPL intensity depend on the applied laser power (\Figref{SI_power}) for different excitation energies. Each of the curves reveals a (quasi) linear increase with laser power, indicating that we do not need to account for multiple-photon processes in the conditions of our experiments. Note that for the laser powers that were used during the experiments we did not observe effects related to heating (i.e., thermal drift in z in constant current/ current in constant height measurements), as can be seen in Fig. 1f where the laser was repeatedly switched on and off during the constant-height measurement. 

\vspace{10pt}
 
\textbf{Efficiency estimation}

Measuring the optical power in front of the UHV viewport allows us to tentatively estimate the number of photons absorbed by a molecule per second \cite{Lombana2020}, $N = \phi \sigma \eta$, where $\phi$ is the photon flux, $\sigma$ the absorption cross-section of the molecule and $\eta$ the enhancement factor, which provides a raw estimate on the photoluminescence and tautomerization yields. Considering the focused laser spot diameter on the sample to be 30 $\mu$m, and the laser power to be 1 $\mu$W we obtain power density of 0.001 $\mu$W $\mu m^{-2}$, which for photons of the energy $h\nu_{in}$ = 1.92 eV corresponds to $\phi$ = 3000 photons nm$^{-2}$ s$^{-1}$. The absorption cross-section of H$_2$Pc, which we estimate from a UV-vis absorption spectrum, is $\sigma$ = 2 $\times$ 10$^{-17}$ cm$^2$. The field enhancement ($E/E_0$) by the picocavity, where $E$ is the field under the tip apex and $E_0$ is the field outside the picocavity, is estimated to be 10$^2$ \cite{Yang2020}, which leads to absorption enhancement of $\eta = (\frac{E}{E_0})^2 = 10^4$. Therefore, we estimate that for the laser power of 1 $\mu$W, the molecule absorbs $N$ = 60000 photons s$^{-1}$. In such excitation conditions and using this simplified calculation, we find that the molecule emits approx. 2000 photons s$^{-1}$ (\Figref{SI_power}b), which leads to a quantum emission yield in the range of 10$^{-2}$, a value that agrees well with works from the literature \cite{Yang2020}. Similarly, we calculate the tautomerization efficiency to be 1 event per 10$^{6}$ absorbed photons.

\vspace{10pt}
 
\textbf{Photoluminescence excitation and action spectroscopy}

The photoluminescence excitation (PLE) and action spectra (AS) are recorded by monitoring, respectively, the spectrally averaged photoluminescence intensity and the tunnelling current as a function of the incoming photon energy. Note that in contrast to earlier STM studies, we use here \textit{optical} action spectroscopy instead of \textit{electron-driven} action spectroscopy \cite{Kim2015}. During the measurement, the tip is positioned at the extremity of a pyrrole subunit of H$_2$Pc in the constant current mode. For stability and drift correction, the "atom tracking" procedure is enabled during the experiment. 

Since it is experimentally challenging to perform resonant TEPL measurements, the detection range of the PLE measurement cannot overlap with the excitation energy. Thus, for high excitation energies $h\nu_{\rm in} >$ 1.87 eV, we monitor the Q$_{x}$ transition intensity (1.800-1.815 eV range), whereas for photon energies resonant with the Q$_{x}$ transition ($h\nu_{\rm in} <$ 1.85 eV) we monitor the intensity of the vibronic peaks (1.642-1.648; 1.665-1.670; 1.710-1.73 eV ranges) that is proportional to the Q$_{x}$ intensity. In these experiments, we use different filters for the detection of the signal, which depend on the excitation range: longpass 633 nm (1.96 eV), longpass 664 nm (1.87 eV) and longpass 700 nm (1.77 eV).

\begin{figure*}[!h]
  \includegraphics{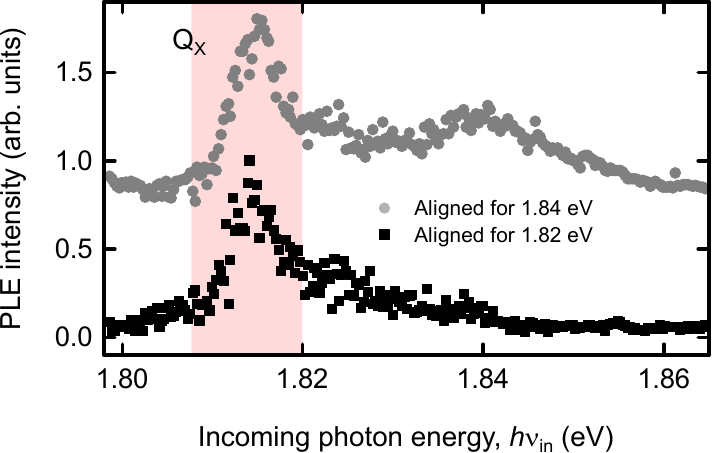}
  \caption{\label{SI_alignment} PLE spectra recorded for two different alignments that are optimized for 1.82 eV and 1.84 eV excitation energy, respectively. The feature that corresponds to the Q$_{x}$ resonance (1.81 eV) is clearly visible on top of the broad alignment distribution that is centred around the optimal excitation wavelength. The traces are vertically offset for clarity. $P$ $<1~\mu$W, $I$ = 10 pA, $V$ = 0.55 V} 
\end{figure*}

The spectra presented in Fig. 2a,b are corrected for the chromatic aberration of the setup. Overall, for such a broad range of excitation energies, the chromatic aberration is sufficient to modify the focusing and thus the light intensity irradiating the tunnelling junction, in particular, when the limits of the energy range are considered. The chromatic aberration is illustrated in \Figref{SI_alignment} where we present two PLE spectra measurements performed on the same molecule in a range of 1.797 - 1.865 eV for two different alignments (optimized for photon energies of 1.82 eV and 1.84 eV, respectively). The feature that corresponds to the Q$_{x}$ resonance (1.81 eV) is clearly visible in the two spectra
demonstrating that the observed resonances originate from the intrinsic properties of the molecule and are not strongly affected by the aberrations of the set-up. 

To reduce the impact of these chromatic aberrations, the PLE experiments (several hours per series) have to be recorded in parts, with an optical realignment step in between each part. The data are acquired such that there is a range of overlapping energies between consecutive spectra, which allows us to compensate for the change in illumination intensity and the impact of chromatic aberration. In the measurement presented in Fig. 2a,b, we used the following partially overlapping ranges: 1.80-1.86 eV; 1.85-1.90 eV; 1.88-1.93 eV; 1.92-2.03 eV. Note that we apply the same scaling factors (1; 3; 4; 5, respectively) for both PLE and action spectra, which also correct the difference between plasmonic enhancements of the used tips. As a consequence of this necessary readjustments and scaling procedures, a comparison of the PLE intensities and tautomerization rates between, for example, Q$_{x}$ and Q$_{y}$ resonances can only be qualitative. Therefore, in our work, we focus on the energy positions of the resonance features rather than on their respective intensities.
\\

\vspace{10pt}
\textbf{Tautomerization analysis}
\begin{figure*}[t!]
  \includegraphics{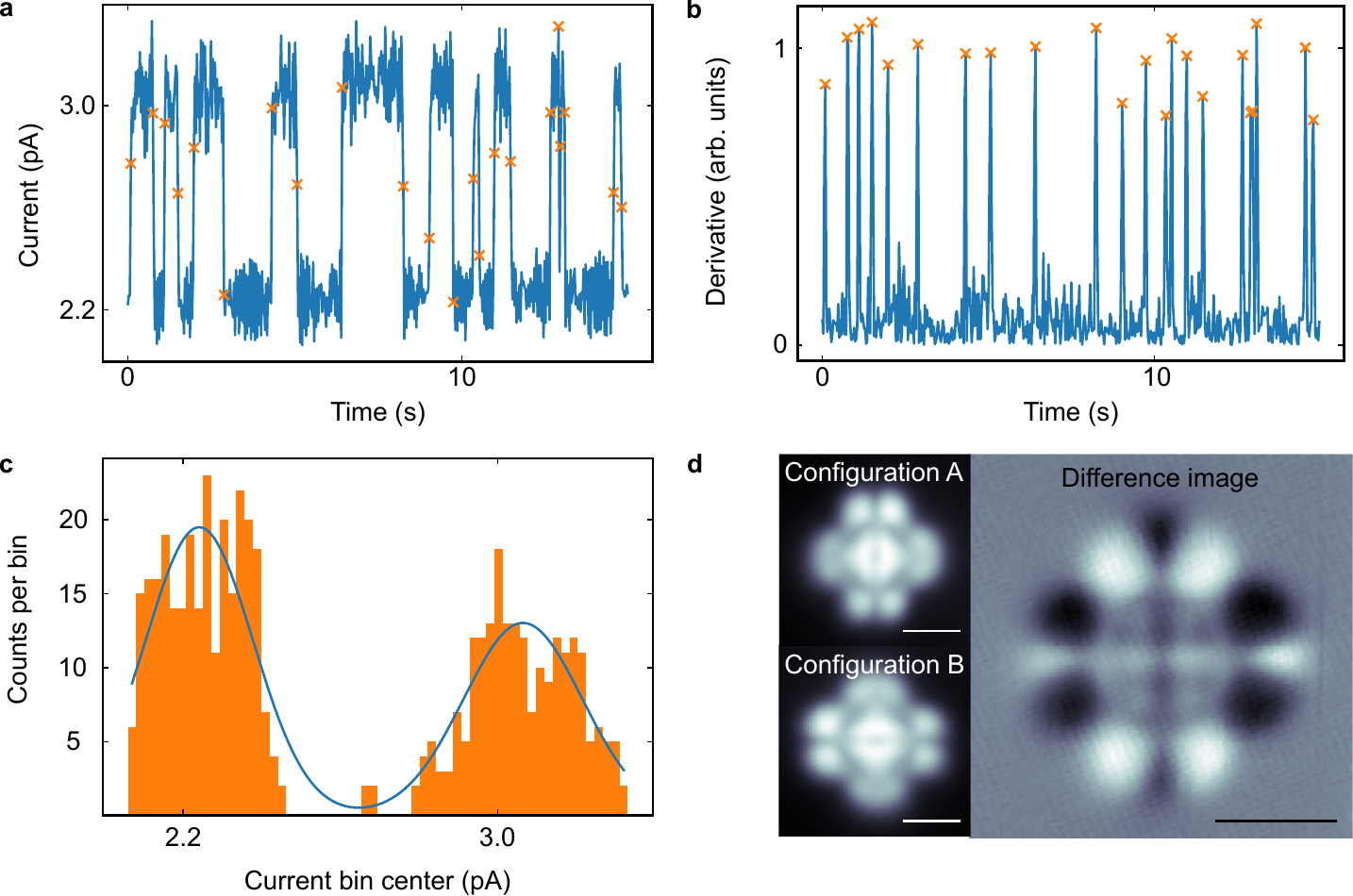}
   \caption{\label{SI_histogram} Tautomerization analysis. a) Current time trace with tautomerization events marked by '$\times$'. b) Numerical derivative of a), c) Histogram of current values in a), the blue line represents a fit, which is a sum of two Gaussians. d) Difference image obtained by subtracting the images representing configuration A and configuration B. Parameters: $V=$ 0.55 V, $I=$ 10 pA, $P\approx$ 15 $\mu$W, $h\nu_{in}=$ 2.00 eV, scale bar: 1 nm.} 
\end{figure*}

In this work, the tautomerization frequency is obtained from the current or vertical (\textit{z}) displacement traces that exhibit two-level switching (\Figref{SI_histogram}a) by calculating a numerical derivative of the signal (\Figref{SI_histogram}b) and counting the number of instances the derivative crosses a certain threshold (marked by '$\times$' in \Figref{SI_histogram}), which is defined by the noise level. We find that in the constant current mode when we monitor the switches in the vertical (\textit{z}) displacement (data presented in Fig. 2 of the main text), the distance change upon tautomerization is about 20 pm. Thus, the error in tautomerization frequency arising for different tip-molecule distances for both tautomers is small and can be neglected. Note that the switching is relatively slow (rates in the Hz range) and we do not observe signatures of faster processes (kHz and more), which would lead to a current distribution (\Figref{SI_histogram}c) dominated by one broad feature due to the tunnelling current averaging. 

The relative configuration population is calculated from a current (or \textit{z} displacement) histogram (\Figref{SI_histogram}c) by fitting two Gaussian distributions and dividing the area under the curve corresponding to the configuration A by the total area under both curves. Assignment of the high and low current states to configurations A and B depends on the tip position on the molecule, thus the local density of states. This is illustrated in \Figref{SI_histogram}d, which is a difference image between configurations A and B that shows in white (black) the areas where the high current corresponds to configuration A (B). Considering the data presented in Fig. 2c of the main manuscript, the tip is in a position where the high current state corresponds to configuration A. In the case of measurements presented in Fig. 4 of the main text, for each position on the molecule, the high current has to be assigned to configuration A or B based on the difference image presented in \Figref{SI_histogram}d. Such configuration identification is possible only in the outer part of the molecule since in the molecular centre the difference in the local density of states is small and features too strongly varying spatial distribution to properly assign the configurations A and B. We remark here that in the case of the tip positions corresponding to the low density of states difference, the noise in the calculated derivative may be on the order of the signal leading to artificially reduced counts (\textit{e.g}. in Fig. 4e).

\vspace{10pt}

\textbf{TEPL background correction}
 
\begin{figure*}[h!]
  \includegraphics{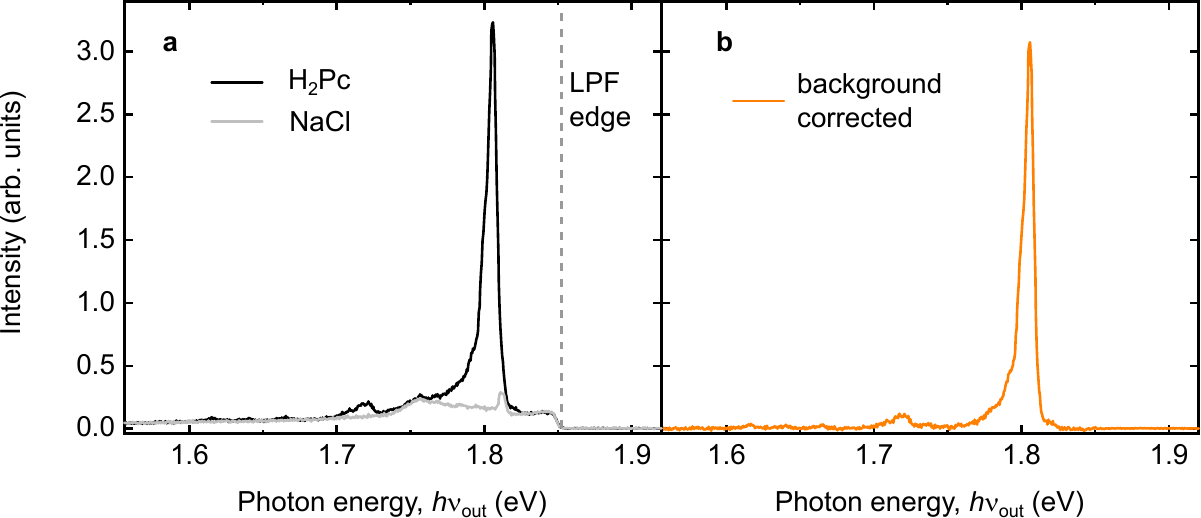}
 \caption{\label{SI_background} a) TEPL spectra recorded on a decoupled H$_{2}$Pc molecule (black line) on NaCl/Ag(111) ($V$ = 0.55 V) and on clean (grey line)  NaCl/Ag(111) ($V$ = - 1 V). $h\nu_{\rm in}$ = 1.953 eV, $P$ $\approx 15~\mu$W, $I$ = 2~pA, $t$ = 90 s. The longpass filter (LPF) edge is indicated. b) Background-corrected TEPL spectrum.} 
\end{figure*}

The TEPL spectra are corrected for the background signal (\Figref{SI_background}) by subtracting a reference spectrum recorded on a clean NaCl surface. We find two main contributions to the background signal (grey trace in \Figref{SI_background}a). First, we observe the broadband signal from the STM tip, which is attributed to the electronic Raman scattering and photoluminescence of the metal \cite{Liu2021b}. Second, we find a spurious sharp emission at $h\nu_{\rm out}$ = 1.82 eV that we assign to the photoluminescence of H$_{2}$Pc molecules deposited on the sample and located outside of the picocavity.
Both signals are independent of the lateral tip position and are detected for tip-sample distances > 1 $\mu$m, thus are taken as a constant background not contributing to the phenomena observed at the sub-molecular scale, which are extremely sensitive to the tip position and the plasmonic enhancement provided by the tip apex (see Fig. 3 of the main text).

\newpage
\clearpage

\subsection{S2 -- Action spectroscopy of HPc$^{-}$}

\begin{figure*}[!h]
  \includegraphics{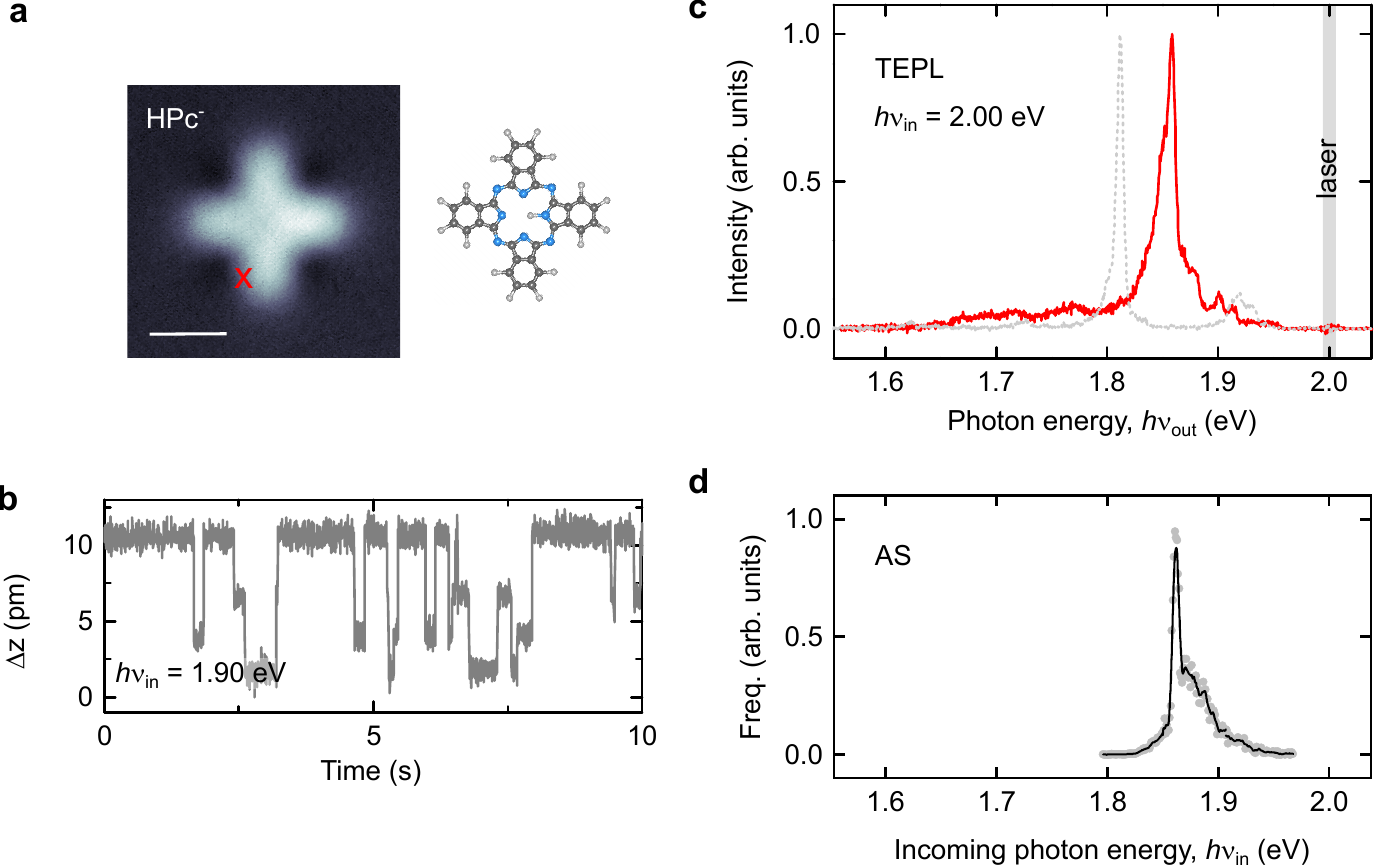}
 \caption{\label{SI_hpc} Tip-enhanced photoluminescence and action spectroscopy of HPc$^{-}$. a) STM image of HPc$^{-}$ $V$ = 0.55 V, $I$ = 2 pA, scale bar 1 nm. The chemical structure of the molecule is presented on the right. b) Tautomerization trace measured with a closed feedback loop; four states corresponding to the four tautomers are observed; $V$ = 0.55 V, $I$ = 10 pA, $h\nu_{\rm in} =$ 1.90 eV, $P~\approx 1~\mu$W. c) TEPL spectrum of HPc$^{-}$ measured at the position indicated in a). $V$ = 0.55 V, $I$ = 2 pA, $h\nu_{\rm in} =$ 2.00~eV, $P~\approx 15~\mu$W, $t$~=~60~s. For reference, the spectrum of H$_{2}$Pc is plotted with a grey dashed line. d) Action spectrum of HPc$^{-}$, $V$ = 0.55 V, $I$ = 10 pA (constant current mode), $P~\approx$ 1 $\mu$W. The measurement was performed using three energy ranges (1.78-1.85; 1.85-1.91; 1.91-1.97 eV)).} 
\end{figure*} 

In \Figref{SI_hpc}, we show that our approach can be used to characterize a different chemical species. We study an HPc$^{-}$ molecule on a NaCl surface (\Figref{SI_hpc}a), which we obtain by removing a central proton of an H$_{2}$Pc. High energy tunnelling electrons \cite{Vasilev2022} or intense laser illumination can be used to deprotonate the molecule. Note that for the measurements on H$_{2}$Pc reported in the main manuscript, relatively low laser powers were used to prevent undesired deprotonation. An STM image of HPc$^{-}$ is presented in \Figref{SI_hpc}a. Its appearance is radically different from an H$_{2}$Pc image recorded at the same bias  (see Fig. 1c of the main manuscript) due to a rigid shift to the higher energy of the molecular electronic levels of HPc$^{-}$ (see \Figref{SI_biasdep}) \cite{Vasilev2022}. In \Figref{SI_hpc}c we show a TEPL spectrum recorded at the position indicated in the STM image. This fluorescence spectrum reveals a resonance at 1.86 eV. In contrast to electroluminescence, which has to be excited indirectly via energy transfer from a neighbouring molecule \cite{Vasilev2022}, photoluminescence of HPc$^{-}$ can be directly obtained. Since HPc$^{-}$ has 4 available hydrogen positions in the central molecular cavity, we observe a 4-state photoinduced ($h\nu_{\rm in} =$ 1.90 eV) switching as presented in \Figref{SI_hpc}b, which we later explore in an action spectroscopy measurement shown in \Figref{SI_hpc}d. Again, we find that phototautomerization is most efficient when the molecule is driven resonantly to the excited state, specific to HPc$^{-}$, showing the versatility of our atomic-scale action spectroscopy approach.

\newpage

\subsection{S3 -- Vibrational features}

\begin{figure*}[!h]
  \includegraphics[width=\linewidth]{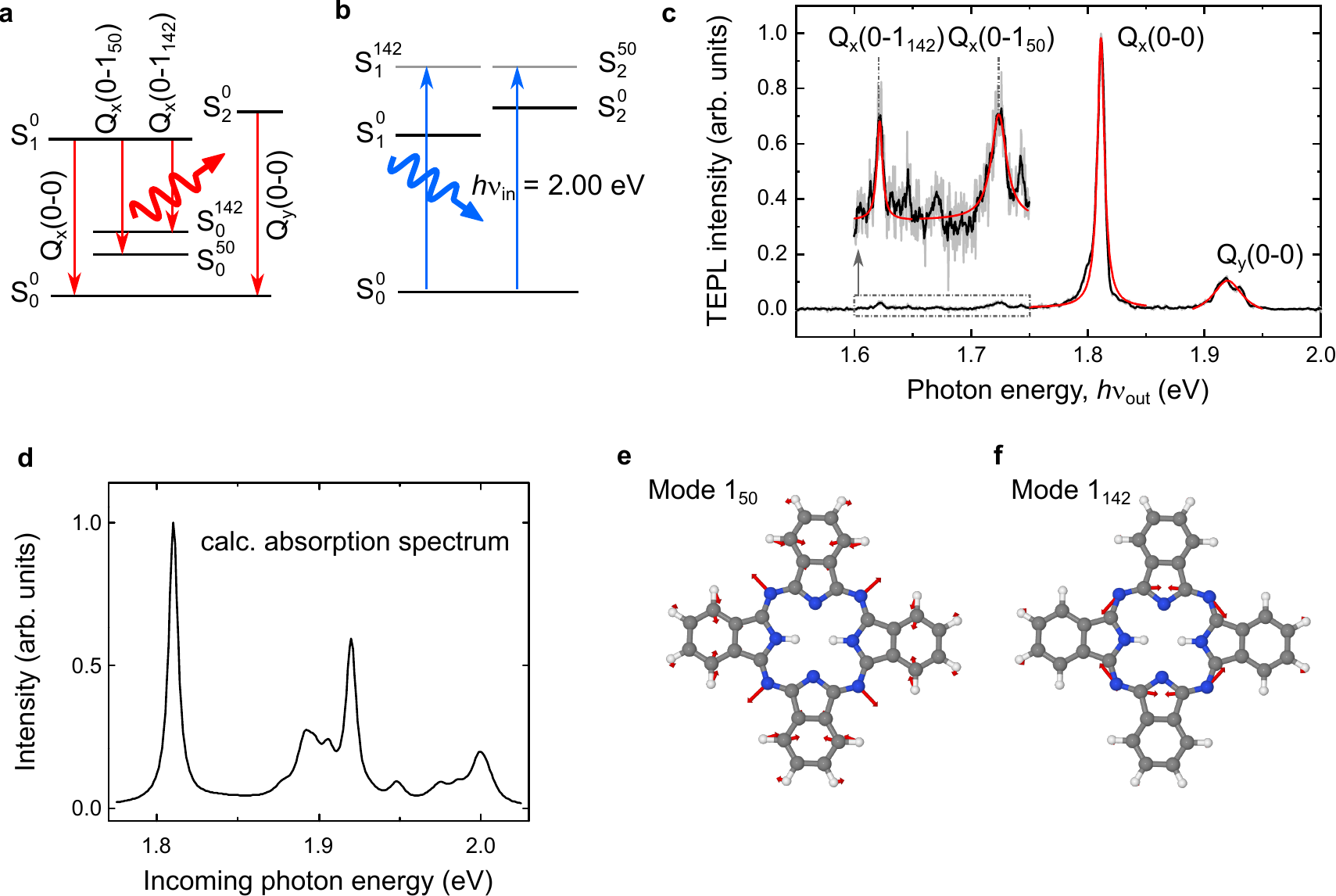}
 \caption{\label{SI_vib} a) Jab\l o\'nski diagram for the emission (the excitation step is not shown). b) Proposed Jab\l o\'nski diagram for $h\nu_{\rm in} =$ 2.00 eV excitation. c) TEPL spectrum of H$_{2}$Pc, $h\nu_{\rm in} =$ 2.00 eV, $t=$~90~s, $V=$ 0.55 V, $I=$ 2 pA, $P \approx$ 15 $\mu$W. The vibrational part of the spectrum (dashed rectangle) is magnified. Red curves show Lorentzian fits used to extract the peak energy positions of the two most intense modes. d) Calculated absorption spectrum of H$_{2}$Pc (see Section S8 for details). e), f) Diagrams of the atom displacements corresponding to the calculated vibrational modes at 656 cm$^{-1}$ (mode 50) and at 1531 cm$^{-1}$ (mode 142) that feature the largest Franck-Condon activity and jointly contribute to the absorption peak observed experimentally around 2 eV.} 
\end{figure*}

In Fig. 2b of the main manuscript, we observe that the excitation at $h\nu_{\rm in} =$ 2.00 eV leads to an increase in the photoluminescence signal. We propose that this effect reflects the coupling of the incoming light to two intense vibrational modes associated to Q$_{x}$ and Q$_{y}$. 

Let us first identify the vibrational modes of interest in the photoluminescence spectrum. \Figref{SI_vib}a and c present, respectively, the emission scheme and a TEPL spectrum of the molecule where the main transitions (Q$_{x}$(0-0) and Q$_{y}$(0-0)) are accompanied by transitions into  vibrational levels of the ground state ($S_0$). Here, we observe two intense modes that we label Q$_{x}$(0-$1_{50}$) and Q$_{x}$(0-$1_{142}$) that appear at $h\nu_{\rm out}$ = 1.724 eV and $h\nu_{\rm out}$ = 1.622 eV, respectively. The numbering originates from the calculations that are discussed below. Taking the energy of Q$_{x}$ as a reference (1.811 eV), we obtain the corresponding energy shifts $E_{S_0^{142}} =$ 0.189 eV (1520 cm$^{-1}$) and $E_{S_0^{50}} =$ 0.087 eV (700 cm$^{-1}$). Note that these peaks are systematically present in Raman and STML spectra of H$_{2}$Pc \cite{Doppagne2020}. 

If we now estimate the energy required to excite the molecule from the ground state $S_0^0$ to the same vibronic modes $142$ and $50$ of, the excited state $S_1$ and $S_2$, respectively, we obtain (see \Figref{SI_vib}b) very similar energies: $S_1^{142}$ = 2.000 eV and $S_2^{50}$ = 2.007 eV (taking $S_2^0$  = 1.920 eV). This indicates an efficient excitation of both $S_{1}$ and $S_{2}$ states for a laser energy $h\nu_{\rm in} =$ 2.00 eV. This explains why at this excitation energy, both TEPL (Fig. 3c) and tautomerization rate (Fig. 3i) maps exhibit homogeneous doughnut-like patterns (\textit{i.e.}, both Q$_x$ and Q$_y$ can be excited).

Our theoretical calculations corroborate this interpretation and allow the assignment of specific modes (see more details in Sections S7 and S8). In \Figref{SI_vib}d we plot the calculated absorption spectrum including both Q$_{x}$ and Q$_{y}$ transitions. We find two most active modes at 656 cm$^{-1}$ (mode 50) and 1531 cm$^{-1}$ (mode 142), which agree well with the values extracted from the experiment. The modes are illustrated in \Figref{SI_vib}e,f. The calculated spectrum has an intense feature nearly 200 meV above the energy of the Q$_{x}$ transition, which agrees very well with the value observed in the PLE experiment (190 meV). The computational details can be found in Section S8.

We remark here that besides these two most intense vibrational modes, the less intense modes also play a role in our measurements. This effect is observed in, for example, Fig. 2c of the main text where the tautomer population features resonances for incoming photon energies 1.87 eV < $h\nu_{in}$ < 1.90 eV and 1.93 eV < $h\nu_{in}$ < 2.00 eV, which are reproduced by our theoretical model.
Note that the origin of the complex spectroscopic fingerprint of the Q$_y$ transition is likely a consequence of vibronic coupling between $S_2$ and vibrational levels of $S_1$\cite{Murray2011}. This feature, however, is beyond the scope of our theoretical modelling that considers the vibronic activity via the Franck-Condon displacements of the respective vibrational modes and neglects effects of non-adiabatic vibronic coupling.

\newpage

\subsection{S4 -- Supplementary TEPL maps}

\begin{figure*}[!h]
  \includegraphics{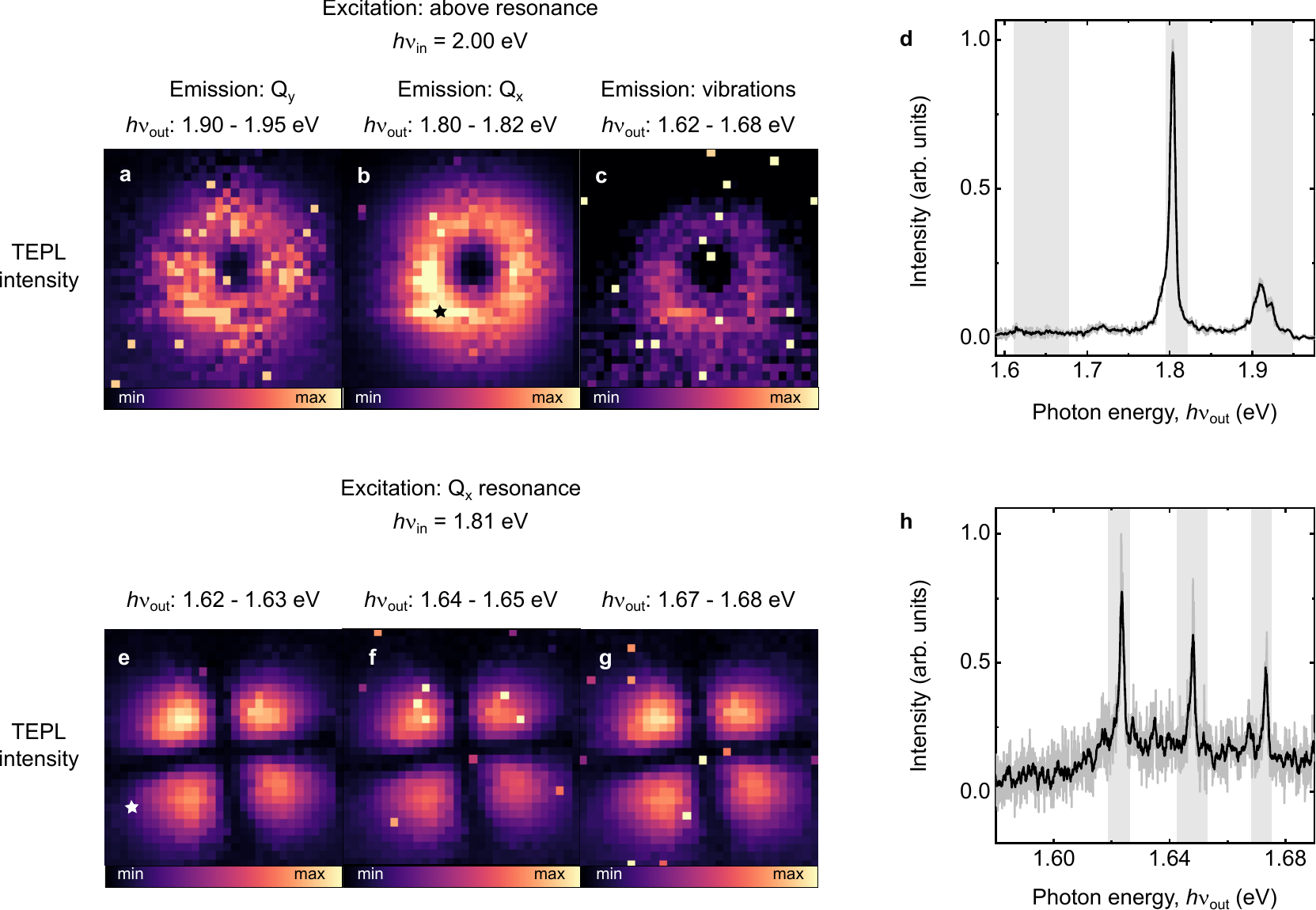}
   \caption{\label{SI_qymap} a-c) TEPL maps of Q$_{y}$ (a), Q$_{x}$ (b) and vibrational (c) emission recorded simultaneously for excitation at $h\nu_{\rm in} =$ 2.00 eV. Size: 4.4 $\times$ 4.4 nm$^{2}$. d) TEPL spectrum recorded at the point marked by a star in b). e-g) TEPL maps of 3 vibrational modes recorded simultaneously at $h\nu_{\rm in} =$ 1.81 eV. h) TEPL spectrum recorded at the point marked by a star in e). The integration ranges used to obtain the TEPL maps are indicated in the TEPL spectra by grey rectangles. $V$ = 0.55 V.} 
\end{figure*}

In \Figref{SI_qymap} we present a comparison between the TEPL maps of different emission lines emitted from one H$_{2}$Pc molecule. In \Figref{SI_qymap}a-c we compare the patterns of Q$_{y}$, Q$_{x}$ and vibrational photoluminescence for a $h\nu_{\rm in} =$ 2.00 eV. We find that the pattern in all three cases takes a doughnut-like shape. Similarly, in \Figref{SI_qymap}e-g we plot the spatial distribution of different vibrational modes acquired for excitation resonant with the Q$_{x}$ mode ($h\nu_{\rm in} =$ 1.81 eV) and find the same patterns for all modes within the probed energy window, but different from the ones recorded for 2.00 eV excitation. Hence, the patterns are determined by the excitation energy and not by the emission energy.

Further analysing the maps in \Figref{SI_qymap}ab, one observes that on top of the doughnut-like pattern, the intensity of neighbouring pixels along the molecular backbone can vary substantially. This effect indicates that the tautomerization occurs at a rate comparable to the acquisition time (15 s per pixel) during which the molecule remains predominantly in one of the two tautomer configurations and no temporal averaging between configurations A and B arises. Note that this effect is not present in the tautomerization map (Fig. 3i) as it occurs in a region where no tunnelling current is detected.

\newpage

\subsection{S5 -- Comparison between STML and TEPL maps}

\begin{figure*}[!h]
  \includegraphics{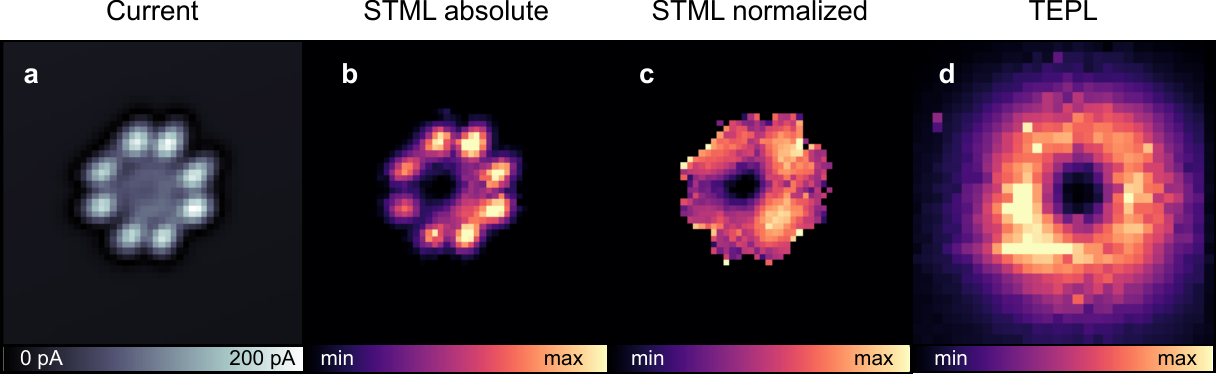}
 \caption{\label{SI_STML} a) Constant height image (current map) of H$_{2}$Pc, $V$ = -2.5 V. b) Simultaneously recorded STML map of the Q$_{x}$ emission (1.78-1.84 eV integration range), 50 $\times$ 50 pixels, $t$ = 0.8 s per pixel. c) Current-normalized STML map. d) Reference TEPL map (from Fig. 3). All images: 4.4 $\times$ 4.4 nm$^{2}$. } 
\end{figure*}

In \Figref{SI_STML} we compare STM-induced luminescence (STML) and TEPL maps acquired on H$_{2}$Pc. These data reveal that TEPL signals can be measured at a larger lateral distance from the centre of the molecule than in STML. Indeed, in the STML experiment (\Figref{SI_STML}b-c), the tunnelling current running through molecular orbitals (\Figref{SI_STML}a) acts as an excitation source. One can therefore not observe any fluorescence for tip positions "aside" from the molecule. In contrast, the TEPL signal relies solely on the tip-plasmon/molecular-exciton coupling, which remains strong even for lateral tip positions where no tunnelling current flows through the molecule \cite{Imada2021}. Therefore, TEPL mapping provides a more complete image of the plasmon-exciton coupling than STML.

Note that the energy degeneracy of Q$_{x}$ and Q$_{y}$ transitions for the two tautomer configurations may be lifted due to the interaction with the substrate, which results in an optical spectrum including two features for each transition, as we have shown previously \cite{Doppagne2020, Roslawska2022}. In this work, we find the transition energies for both tautomers to be degenerate and, as a consequence, all TEPL and STML maps presented in this work exhibit (at least) four-fold symmetry. This effect is also responsible for the contrast present in \Figref{SI_STML}c where the two transition dipoles cannot be identified and are spatially averaged.

\newpage

\subsection{S6 -- Supplementary discussion on the patterns in TEPL and tautomerization maps}

\begin{figure*}[!h]
  \includegraphics[width=\linewidth]{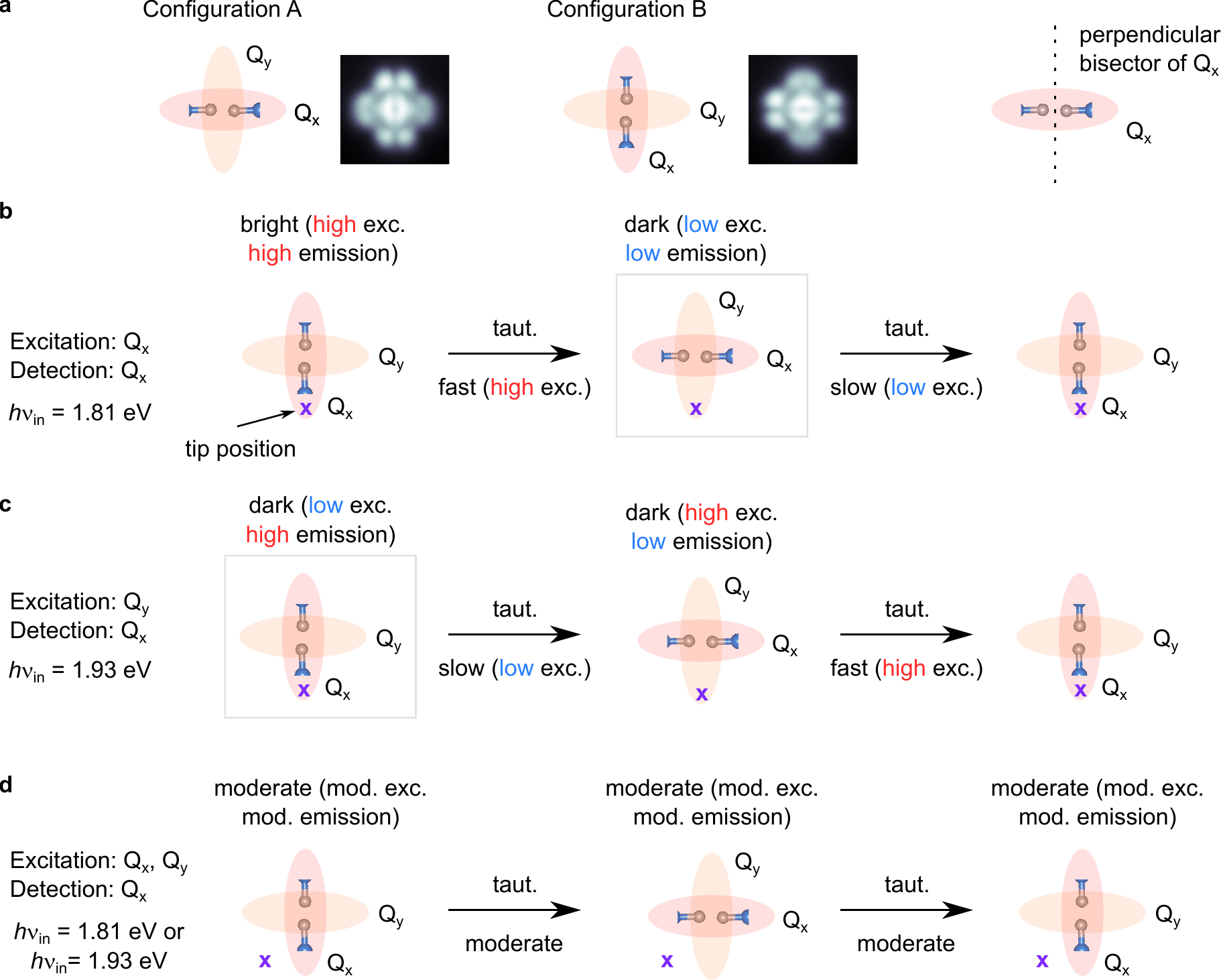}
 \caption{\label{SI_scheme} Scheme presenting the result of an interplay between optical properties and tautomerization (taut.) of H$_{2}$Pc. For simplicity, only the two hydrogens that tautomerize are shown. The grey boxes indicate the configuration in which the molecule is temporarily locked.} 
\end{figure*}

In \Figref{SI_scheme} we present an intuitive understanding of the TEPL and tautomerization rate maps recorded at resonance conditions. The summary of the orientation of dipoles for both configurations A and B is shown in \Figref{SI_scheme}a. \Figref{SI_scheme}b shows the mechanism of the tautomerization evolution for the excitation resonant with the Q$_{x}$ dipole and an initial tip position (cross in \Figref{SI_scheme}b-d)) at the extremity of the Q$_{x}$ dipole. In this case, both the excitation and emission of Q$_{x}$ are efficient due to the high coupling to the picocavity plasmon, therefore the residence time of the molecule in the initial tautomer is short. Once tautomerization occurs, the tip is now located at the perpendicular bisector of Q$_{x}$ (see the sketch in \Figref{SI_scheme}a) that leads to low coupling with the picocavity and inefficient excitation and emission of Q$_{x}$. This, in turn, results in a long residence time, \textit{i. e.} the molecule is temporarily locked in this configuration (grey box). Only once the tautomerization occurs the cycle can repeat.

\Figref{SI_scheme}c shows a similar scheme considering the excitation of the Q$_{y}$ dipole and emission of Q$_{x}$, following the same approach as in the experiment. Taking the same initial conditions as in \Figref{SI_scheme}b, the tip is located at the perpendicular bisector of Q$_{y}$, which leads to inefficient excitation, thus long residence time and temporary configuration locking. Note that another consequence of low excitation is the fact that the fluorescence from this tautomer configuration is not observed, even though this configuration favours radiative decay of Q$_{x}$. After switching, the tip is located at the extremity of the Q$_{y}$ dipole. This configuration results in efficient excitation, however, the emission (from Q$_{x}$) is suppressed and the configuration is dark. Since the tautomerization is driven by excitation, rather than emission, the switch from that configuration is fast. Taking the two cases together, we see that once the tip is located at the perpendicular bisector of the dipole, with which the excitation is resonant, the molecule is locked in a tautomeric configuration which leads to suppressed emission.

Finally, in \Figref{SI_scheme}d we present the evolution of configurations for a tip position at the diagonal between the Q$_{x}$ and Q$_{y}$ dipoles. In this case, for both tautomers, the excitation and emission probabilities are moderate. This results in a relatively high switching frequency (no configuration locking) and an efficient TEPL signal generation.
\newpage

\newpage

\subsection{S7 -- Theoretical model}

Here we provide the essential information about the theoretical model that we use to reproduce the experimentally obtained TEPL maps and maps of tautomerization rates. We outline the technical details of the model and discuss the parameters that we use to compare the calculations with the experimental results. 

The molecule quantum state can be described as a superposition of basis functions $|\psi\rangle$ constructed as the direct product of the tautomer configurations, the electronic state of the molecule, and the vibrational state as $|\psi\rangle = |\psi_{\rm T}\rangle \otimes |\psi_{\rm E}\rangle\otimes|\psi_{\rm vib}\rangle$. We express the tautomer configuration $|\psi_{\rm T}\rangle$ in the basis $\{ |{\rm A}\rangle, |{\rm B}\rangle \}$ (tautomer A and B configuration, respectively). The molecule electronic state $|\psi_{\rm E}\rangle$ is approximated in the basis $\{  |{\rm G}\rangle, |{\rm X}\rangle, |{\rm Y}\rangle  \} $, where $|{\rm G}\rangle$ is the electronic ground state, and $|{\rm X}\rangle$ and $|{\rm Y}\rangle$ are singlet excited states whose transition dipole moment is oriented along axes $X$ and $Y$. We choose the $(X, Y)$ coordinate system to be rigidly linked with the substrate, as opposed to the coordinate system $(x, y)$ where the $x$ axis is chosen to be along the pair of central hydrogens and $y$ is perpendicular to it. The vibrational state $|\psi_{\rm vib}\rangle$ is described in the basis of the vibrational ground state $|0\rangle$ and the singly vibrationally excited states,$|1_m\rangle$. Each of these states $|1_m\rangle$ contains one quantum of a vibrational mode $m$ and all other modes in the vibrational ground state. Restricting our description to single vibrational excitations is justified by the rigidity of the H$_2$Pc molecule resulting in less pronounced vibronic spectral features. We include in the model only the vibrational modes with the strongest Franck-Condon activity, as discussed in Section S8. These states can appear in the absorption spectrum as satellites of Q$_x$ and Q$_y$ at a higher energy.
In this basis, we further define operators $\tau_{IJ}\equiv |I\rangle\langle J|\otimes I_{\rm E}\otimes I_{\rm vib}$, and $\sigma_{IJ}\equiv I_{\rm T}\otimes |I\rangle\langle J| \otimes I_{\rm vib} $,  where $I_{\rm T}$, $I_{\rm E}$ and $I_{\rm vib}$ are the identity operator acting on the molecular tautomar, electronic, and vibrational subspace, respectively. Using this notation, one can express the Hamiltonian of the pumped molecule (using the rotating-wave approximation and transforming the Hamiltonian into the rotating frame) as:
\begin{align}
    H=&\tau_{\rm AA}\bigg\{(E_x-\omega)\sigma_{\rm XX}+(E_y-\omega)\sigma_{\rm YY}\nonumber\\
    +&\Omega_{\rm X}^{\rm A} (\sigma_{\rm GX}+\sigma_{\rm XG})+\Omega_{\rm Y}^{\rm A} (\sigma_{\rm GY}+\sigma_{\rm YG})\nonumber\\
    +&\sum_m d_m^{x}\Omega_{\rm X}^{\rm A} \left[\sigma_{{\rm GX}}(b_m-b_m^\dagger)+\sigma_{{\rm X}{\rm G}}(b_m^\dagger-b_m)\right]\nonumber\\
    +&\sum_m d_m^{y}\Omega_{\rm Y}^{\rm A} \left[\sigma_{{\rm GY}}(b_m-b_m^\dagger)+\sigma_{{\rm Y}{\rm G}}(b_m^\dagger-b_m)\right]\bigg\}\nonumber\\
    +&\tau_{\rm BB}\bigg\{(E_y-\omega)\sigma_{\rm XX}+(E_x-\omega)\sigma_{\rm YY}\nonumber\\
    +&\Omega_{\rm X}^{\rm B} (\sigma_{\rm GX}+\sigma_{\rm XG})+\Omega_{\rm Y}^{\rm B} (\sigma_{\rm GY}+\sigma_{\rm YG})\nonumber\\
    +&\sum_m d_m^{y}\Omega_{\rm X}^{\rm B} \left[\sigma_{{\rm GX}}(b_m-b_m^\dagger)+\sigma_{{\rm X}{\rm G}}(b_m^\dagger-b_m)\right]\nonumber\\
    +&\sum_m d_m^{x}\Omega_{\rm Y}^{\rm B} \left[\sigma_{{\rm GY}}(b_m-b_m^\dagger)+\sigma_{{\rm Y}{\rm G}}(b_m^\dagger-b_m)\right]\bigg\}\nonumber\\
    +&\sigma_{\rm GG}\nu^{\rm G}_m b^\dagger_m b_m+\tau_{\rm AA}(\sigma_{\rm XX}\nu^{x}_m b^\dagger_m b_m+\sigma_{\rm YY}\nu^{y}_m b^\dagger_m b_m)+\tau_{\rm BB}(\sigma_{\rm XX}\nu^{y}_m b^\dagger_m b_m+\sigma_{\rm YY}\nu^{x}_m b^\dagger_m b_m)\label{eq:Hamiltonian}
\end{align}
Here $E_x=1.81$\,eV ($E_y=1.92$\,eV) is the energy of the Q$_x$ (Q$_y$) transition, $\Omega_{I}^{J}$ (with $I\in\{{\rm X},{\rm Y}\}$ and $J\in\{{\rm A},{\rm B}\}$) is the coherent pumping amplitude of the $X$ ($Y$) oriented excitation in the A (B) tautomer configuration. $\nu^{j}_m$ are the energies of the Franck-Condon active vibrations evaluated in the electronic state $j\in\{{\rm G}, x, y\}$ that jointly contribute to the vibronic absorption peaks in the spectrum. The vibrations are represented in the Hamiltonian by their bosonic creation and annihilation operators $b_m^\dagger$ and $b_m$ for mode $m$. The first line of the conditional Hamiltonian in Eq.\,\eqref{eq:Hamiltonian} represents the exciton energies conditioned on the tautomer configuration of the molecule. This is the key ingredient that allows us to describe how molecular tautomerization can be selectively driven by external illumination. The coherent pumping amplitudes $\Omega_{I}^J$ are considered as dependent on the tip position and are closely related to the plasmon-induced decay rates $\kappa_I^J$ (discussed below) as $\Omega_{I}^J=\Omega_0\sqrt{\kappa_I^J/\gamma_{\rm GS}}$, with $\Omega_0$ being a pumping amplitude, and $\gamma_{\rm GS}$ being a parameter setting the plasmon-enhanced decay rate. Finally, we consider that the interaction of light with the vibronic states is 
governed by the Franck-Condon principle. We, therefore, introduce the parameter $d_m^{j}$ of mode $m$ in the exciton $j\in \{ x, y \}$, which is a dimensionless scaling factor (related to the Franck-Condon coefficient of the respective vibration) adjusting the light-matter interaction strength in the model. We obtain the $d_m^{j}$ parameters of the H$_2$Pc molecule using density-functional theory (DFT) and time-dependent DFT (TDDFT) as discussed in Section S8.

Besides the Hamiltonian, the dynamics of the molecular states are also driven by incoherent processes. We represent these processes in the density-matrix formalism using the Lindblad superoperators defined for a generic operator $O$ and a generic decay rate $\gamma$ as:
\begin{align}
    \gamma \mathcal{L}(O)=\gamma\left( O\rho O^\dagger-\frac{1}{2}\{ O^\dagger O,\rho \} \right),
\end{align}
where $\rho$ is the density matrix of the system and $\dagger$ denotes the Hermitian conjugate. In particular, we define the dynamics of the density matrix as follows: 
\begin{align}
    \frac{{\rm d} \rho}{{\rm d}t}=&i[\rho, H]\nonumber\\
    +&\gamma_{\rm GX}^{\rm A}\mathcal{L}(\tau_{\rm AA}\sigma_{\rm GX})+\gamma_{\rm GX}^{\rm B}\mathcal{L}(\tau_{\rm BB}\sigma_{\rm GX})+\gamma_{\rm GY}^{\rm A}\mathcal{L}(\tau_{\rm AA}\sigma_{\rm GY})+\gamma_{\rm GY}^{\rm B}\mathcal{L}(\tau_{\rm BB}\sigma_{\rm GY})\nonumber\\
    +&\sum_m\gamma_{\rm SS}\mathcal{L}(b_m)\nonumber\\
    +&\sum_m \gamma_{{\rm GX}_m}^{\rm A}\mathcal{L}(\tau_{\rm AA}\sigma_{{\rm GX}}b_m)+\gamma_{{\rm GX}_m}^{\rm B}\mathcal{L}(\tau_{\rm BB}\sigma_{{\rm GX}}b_m)+\gamma_{{\rm GY}_m}^{\rm A}\mathcal{L}(\tau_{\rm AA}\sigma_{{\rm GY}}b_m)+\gamma_{{\rm GY}_m}^{\rm B}\mathcal{L}(\tau_{\rm BB}\sigma_{\rm GY}b_m)\nonumber\\
    +&\sum_m \gamma_{{\rm GX}_m}^{\rm A}\mathcal{L}(\tau_{\rm AA}\sigma_{{\rm GX}}b_m^\dagger)+\gamma_{{\rm GX}_m}^{\rm B}\mathcal{L}(\tau_{\rm BB}\sigma_{{\rm GX}}b_m^\dagger)+\gamma_{{\rm GY}_m}^{\rm A}\mathcal{L}(\tau_{\rm AA}\sigma_{{\rm GY}}b_m^\dagger)+\gamma_{{\rm GY}_m}^{\rm B}\mathcal{L}(\tau_{\rm BB}\sigma_{\rm GY}b_m^\dagger)\nonumber\\
    +&\gamma_{\rm AB}^{\rm XA}\mathcal{L}(\sigma_{\rm XX}\tau_{\rm BA})+\gamma_{\rm AB}^{\rm YA}\mathcal{L}(\sigma_{\rm YY}\tau_{\rm BA})+\gamma_{\rm AB}^{\rm XB}\mathcal{L}(\sigma_{\rm XX}\tau_{\rm AB})+\gamma_{\rm AB}^{\rm YB}\mathcal{L}(\sigma_{\rm YY}\tau_{\rm AB})
    \nonumber\\
    +&\gamma_{xy}[\mathcal{L}(\tau_{\rm AA}\sigma_{\rm XY})+\mathcal{L}(\tau_{\rm BB}\sigma_{\rm YX})]\nonumber\\
    +&\gamma_{\phi}[\mathcal{L}(\sigma_{\rm XX})+\mathcal{L}(\sigma_{\rm YY})].\label{eq:master}
\end{align}
The first line is the commutator of the density matrix with the Hamiltonian and represents the coherent part of the dynamics. The second line of Eq.\,\eqref{eq:master} represents the decay of the singlet excited state $|{\rm X}\rangle$ ($|{\rm Y}\rangle$) into the ground state with rates $\gamma_{\rm GX}^{\rm A}$ and $\gamma_{\rm GX}^{\rm B}$ ($\gamma_{\rm GY}^{\rm A}$, $\gamma_{\rm GY}^{\rm B}$) dependent on the position of the tip and the tautomer configuration:
\begin{align}
    \gamma_{\rm GX}^{\rm A}({\bf r}_{\rm tip})&=\kappa_{\rm GX}^{\rm A}({\bf r}_{\rm tip})+\gamma_{\rm GS}^0,\\
    \gamma_{\rm GY}^{\rm A}({\bf r}_{\rm tip})&=\kappa_{\rm GY}^{\rm A}({\bf r}_{\rm tip})+\gamma_{\rm GS}^0,\\
    \gamma_{\rm GX}^{\rm B}({\bf r}_{\rm tip})&=\kappa_{\rm GX}^{\rm B}({\bf r}_{\rm tip})+\gamma_{\rm GS}^0,\\
    \gamma_{\rm GY}^{\rm B}({\bf r}_{\rm tip})&=\kappa_{\rm GY}^{\rm B}({\bf r}_{\rm tip})+\gamma_{\rm GS}^0,
\end{align}
with
\begin{align}
    \kappa_{\rm GX}^{\rm A}({\bf r}_{\rm tip})&=\frac{|g_{\rm X}^{\rm A}({\bf r}_{\rm tip})|^2}{\max(|g_{\rm X}^{\rm A}({\bf r}_{\rm tip})|^2)}\gamma_{\rm GS},\\
    \kappa_{\rm GY}^{\rm A}({\bf r}_{\rm tip})&=\frac{|g_{\rm Y}^{\rm A}({\bf r}_{\rm tip})|^2}{\max(|g_{\rm X}^{\rm A}({\bf r}_{\rm tip})|^2)}\gamma_{\rm GS},\\
    \kappa_{\rm GX}^{\rm B}({\bf r}_{\rm tip})&=\frac{|g_{\rm X}^{\rm B}({\bf r}_{\rm tip})|^2}{\max(|g_{\rm X}^{\rm A}({\bf r}_{\rm tip})|^2)}\gamma_{\rm GS},\\
    \kappa_{\rm GY}^{\rm B}({\bf r}_{\rm tip})&=\frac{|g_{\rm Y}^{\rm B}({\bf r}_{\rm tip})|^2}{\max(|g_{\rm X}^{\rm A}({\bf r}_{\rm tip})|^2)}\gamma_{\rm GS},
\end{align}
where $\gamma_{\rm GS}$ and $\gamma_{\rm GS}^0$ are parameters. The former accounts for the plasmon-induced decay due to the coupling of the excitons to the picocavity, and the latter accounts for additional decay processes (for example due to the excitation of electron-hole pairs in the nearby metal surface) that do not depend on the position of the tip. $g_{I}^{J}$ ($I\in\{{\rm X}, {\rm Y}\}$ and $J\in\{{\rm A},{\rm B}\}$) are plasmon-exciton coupling strengths calculated as the overlap of the respective transition density $\rho_I^J$ and the plasmonic quasi-static potential $\phi_{\rm pl}$ \cite{Roslawska2022}:
\begin{align}
    g_I^J({\bf r}_{\rm tip})=\int\rho_I^J({\bf r})\phi_{\rm pl}({\bf r}-{\bf r}_{\rm tip})\,{\rm d}{\bf r}. 
\end{align}
Here ${\bf r}_{\rm tip}$ is the position of the tip. The potential $\phi_{\rm pl}$ is approximated as a potential generated by a pair of point charges of opposite magnitude, one positioned at $z$ = 1.1\,nm ($x=x_{\rm tip}$, $y=y_{\rm tip}$) above the plane of the molecule ($z$ = 0) and another positioned at $z = $-2.1\,nm ($x=x_{\rm tip}$, $y=y_{\rm tip}$) below this plane. Such a potential reproduces well the localization of the plasmonic electric field found in plasmonic picocavities \cite{Benz2016, Roslawska2022} and is a computationally inexpensive alternative to the potentials obtained as a computationally more demanding full solution of Poisson's equation in plasmonic structures. The transition densities, generalizing the transition-dipole moment of the molecule, are calculated using TDDFT (see Section S8).  

In the third line of Eq.\,\eqref{eq:master}, the vibronically excited states are assumed to decay to the vibrational ground state of the respective singlet excitation with rate $\gamma_{\rm SS}$. The fourth and fifth line represent the decay from the excited states into the electronic ground state of the molecule while annihilating (creating) a vibrational quantum with rates 
\begin{align}
    \gamma_{{\rm G}X_m}^{\rm A}={d_{m}^{x}}^2(\kappa_{{\rm GX}}^{\rm A}+\gamma_{\rm GS}^0),\\
    \gamma_{{\rm G}Y_m}^{\rm A}={d_{m}^{y}}^2(\kappa_{{\rm GX}}^{\rm B}+\gamma_{\rm GS}^0),\\
    \gamma_{{\rm G}X_m}^{\rm B}={d_{m}^{y}}^2(\kappa_{{\rm GX}}^{\rm B}+\gamma_{\rm GS}^0),\\
    \gamma_{{\rm G}Y_m}^{\rm B}={d_{m}^{x}}^2(\kappa_{{\rm GY}}^{\rm B}+\gamma_{\rm GS}^0).
\end{align}

The sixth line of Eq.\,\eqref{eq:master} represents the tautomerization process with the rates $\gamma_{\rm AB}^{IJ}$ ($I\in\{{\rm X}, {\rm Y}\}$ and $J\in\{{\rm A}, {\rm B}\}$) that depend on the particular exciton X (Y) and tautomer configuration A (B). We discuss the choice of values for the tautomerization rates $\gamma_{\rm AB}^{IJ}$ below. In our model, tautomerization only occurs when the molecule is excited, which is achieved by including the respective operators ($\sigma_{\rm XX}$ and $\sigma_{\rm YY}$) in the definition of the Lindblad superoperators.

The second last line of Eq.\,\eqref{eq:master} represents the decay of the singlet exciton of higher energy to the singlet exciton of lower energy with rate $\gamma_{xy}$ (\textit{e.g.} mediated by molecular vibrations). This process is dependent on the tautomer configuration which is made explicit by including the operators $\tau_{\rm AA}$ ($\tau_{\rm BB}$) in the definition. 

And finally, the last line of  Eq.\,\eqref{eq:master} introduces the dephasing processes with dephasing rate $\gamma_\phi$. Dephasing generally does not influence the decay rate of the states, but introduces additional spectral broadening of the respective resonances that could emerge from the interaction of the molecule with its environment.

We derive the rate equations governing the tautomerization and photon emission from the master equation [Eq.\,\eqref{eq:master}] assuming that the molecule is weakly excited. We first obtain the equations of motion for the mean value $\langle O\rangle = {\rm Tr}\{ O \rho \}$ of a generic operator $O$ following the prescription:
\begin{align}
    \dot{\langle O \rangle}=\frac{1}{i\hbar}\left\langle  [O, H]  \right\rangle + \sum_j \left\langle \gamma_j \left(A_j^\dagger O A_j - \frac{1}{2}\left\{ A_j^\dagger A_j, O \right\} \right) \right\rangle,
\end{align}
where the last term represents the sum over all Lindblad terms introduced in Eq.\,\eqref{eq:master}, where the respective operators enter the Lindblad terms as $\mathcal{L}(A_j)$, and $\gamma_j$ represent the respective decay rates. 
Using this general expression, it is possible to derive a hierarchy of equations for the mean values of operators. This hierarchy of equations can be solved for the steady state where all time derivatives are equal to zero.
For example, the equations for the tautomerization rates follow from:
\begin{align}
    \dot{\langle \tau_{\rm AA} \rangle}&=\gamma_{\rm AB}^{\rm XB}\langle \sigma_{\rm XX}\tau_{\rm BB} \rangle+\gamma_{\rm AB}^{\rm YB}\langle \sigma_{\rm YY}\tau_{\rm BB} \rangle-\gamma_{\rm AB}^{\rm XA}\langle \sigma_{\rm XX}\tau_{\rm AA} \rangle-\gamma_{\rm AB}^{\rm YA}\langle\sigma_{\rm YY} \tau_{\rm AA} \rangle,\\
    \dot{\langle \tau_{\rm BB} \rangle}&=\gamma_{\rm AB}^{\rm XA}\langle \sigma_{\rm XX}\tau_{\rm AA} \rangle+\gamma_{\rm AB}^{\rm YA}\langle \sigma_{\rm YY}\tau_{\rm AA} \rangle-\gamma_{\rm AB}^{\rm XB}\langle \sigma_{\rm XX}\tau_{\rm BB} \rangle-\gamma_{\rm AB}^{\rm YB}\langle\sigma_{\rm YY} \tau_{\rm BB} \rangle.
\end{align}
We next define:
\begin{align}
    N_{\rm X}^{\rm A}N_{\rm A}\approx\langle \sigma_{\rm XX}\tau_{\rm AA}\rangle,\label{seq:fact1}\\
    N_{\rm X}^{\rm B}N_{\rm B}\approx\langle \sigma_{\rm XX}\tau_{\rm BB}\rangle,\\
    N_{\rm Y}^{\rm A}N_{\rm A}\approx\langle \sigma_{\rm YY}\tau_{\rm AA}\rangle,\\
    N_{\rm Y}^{\rm B}N_{\rm B}\approx\langle \sigma_{\rm YY}\tau_{\rm BB}\rangle,\label{seq:fact4}
\end{align}
where the tautomer populations are:
\begin{align}
    N_{\rm A}=\langle \tau_{\rm AA}\rangle, \\
    N_{\rm B}=\langle \tau_{\rm BB}\rangle,
\end{align}
and $N_{\rm X}^{\rm A}$, $N_{\rm X}^{\rm B}$, $N_{\rm Y}^{\rm A}$, and $N_{\rm Y}^{\rm B}$ are the conditional populations of the ${ X}$ (${Y}$) oriented excitation assuming that the tautomer is in the ${\rm A}$ or ${\rm B}$ configuration. We note that the factorisation suggested in Eq.\,\eqref{seq:fact1} to Eq.\,\eqref{seq:fact4} is a good approximation due to the separation of time scales present in the system (fast optical excitation and exciton decay versus slow tautomerization). 
We next manipulate the equations and derive the expressions for the photon emission rates and tautomerization rates. The tautomerization dynamics in the H$_2$Pc molecule can be described with the help of the rate equations:
\begin{align}
    \frac{{\rm d}N_{\rm B}}{{\rm d}t}&=\Gamma_{\rm A}N_{\rm A}-\Gamma_{\rm B}N_{\rm B},\label{eq:rate1}\\
    \frac{{\rm d}N_{\rm A}}{{\rm d}t}&=-\Gamma_{\rm A}N_{\rm A}+\Gamma_{\rm B}N_{\rm B},\label{eq:rate2}
\end{align}
where the rates $\Gamma_{\rm A}$ and $\Gamma_{\rm B}$, defining the tautomerization rate $\Gamma={\rm Tr}\{\tau_{\rm AA}[\gamma_{\rm BA}^{\rm XA}\mathcal{L}(\sigma_{\rm XX}\tau_{\rm BA})+\gamma_{\rm BA}^{\rm YA}\mathcal{L}(\sigma_{\rm YY}\tau_{\rm BA})]\}\approx\Gamma_{\rm A}N_{\rm A}=\Gamma_{\rm B}N_{\rm B}$ shown in Fig.\,\ref{fig3}j-o of the main text, depend on the excitation conditions of the molecular electronic states as:
\begin{align}
    \Gamma_{\rm A}= \gamma_{\rm AB}^{\rm XA}N_{\rm X}^{\rm A}+\gamma_{\rm AB}^{\rm YA}N_{\rm Y}^{\rm A},\\
    \Gamma_{\rm B}=\gamma_{\rm AB}^{\rm XB}N_{\rm X}^{\rm B}+\gamma_{\rm AB}^{\rm YB}N_{\rm Y}^{\rm B},
\end{align}
The rate equations Eq.\,\eqref{eq:rate1} and Eq.\,\eqref{eq:rate2} result in the following mean populations of the two tautomer configurations:
\begin{align}
    N_{\rm B} = \frac{\gamma_{\rm AB}^{\rm XA}N_{\rm X}^{\rm A}+\gamma_{\rm AB}^{\rm YA}N_{\rm Y}^{\rm A}}{\gamma_{\rm AB}^{\rm XA}N_{\rm X}^{\rm A}+\gamma_{\rm AB}^{\rm YA}N_{\rm Y}^{\rm A}+\gamma_{\rm AB}^{\rm XB}N_{\rm X}^{\rm B}+\gamma_{\rm AB}^{\rm YB}N_{\rm Y}^{\rm B}},\label{eq:NL}\\
    N_{\rm A} = \frac{\gamma_{\rm AB}^{\rm XB}N_{\rm X}^{\rm B}+\gamma_{\rm AB}^{\rm YB}N_{\rm Y}^{\rm B}}{\gamma_{\rm AB}^{\rm XA}N_{\rm X}^{\rm A}+\gamma_{\rm AB}^{\rm YA}N_{\rm Y}^{\rm A}+\gamma_{\rm AB}^{\rm XB}N_{\rm X}^{\rm B}+\gamma_{\rm AB}^{\rm YB}N_{\rm Y}^{\rm B}}.\label{eq:NH}
\end{align}
The conditional populations can be obtained as a function of the excitation frequency $\omega$ and tip position ${\bf r}_{\rm tip}$ by e.g. numerically solving the full quantum Master equation [Eq.\,\eqref{eq:master}]. In this paper we implement the full numerical solution of Eq.\,\eqref{eq:master} using methods described in \cite{neuman2019raman}. 

Finally, the TEPL intensity $\mathcal I$, shown as the theoretically calculated photon maps in Fig.\,\ref{fig3}d-f of the main text, emitted by the decaying Q$_x$ is calculated as:
\begin{align}
    \mathcal{I}({\bf r}_{\rm tip}) = N_{\rm X}^{\rm A}({\bf r}_{\rm tip})N_{\rm A}({\bf r}_{\rm tip}) \kappa_{\rm X}^{\rm A}({\bf r}_{\rm tip})+N_{\rm Y}^{\rm B}({\bf r}_{\rm tip})N_{\rm B}({\bf r}_{\rm tip}) \kappa_{\rm Y}^{\rm B}({\bf r}_{\rm tip}).\label{eq:IQx}
\end{align}
This expression represents the fact that the photon emission happens only through the radiative plasmonic channel via the decay rate $\kappa_{\rm X}^{\rm A}$ ($\kappa_{\rm Y}^{\rm B}$). The photon emission probability is further dependent on the probability to populate a particular tautomer configuration ($N_{\rm A}$ and $N_{\rm B}$) and to populate the Q$_x$ exciton in the respective tautomer configuration ($N_{\rm X}^{\rm A}$ and $N_{\rm Y}^{\rm B}$). The combination of the spatial dependence of all of these ingredients is finally reflected in the observed photon maps and can result in non-trivial nuances of the spatial distribution of the collected Q$_x$ light intensity under different illumination conditions. 

\textbf{Values of parameters used in the theoretical model}
\newline To reproduce the experimental photon maps we fix the parameters of the model (see Table\,\ref{tab:param}). The choice of particular model parameters is largely guided by previous experimental observations and is therefore not arbitrary. For example, the plasmon-enhanced excitonic decay rate has been calculated and experimentally observed to be $\sim 1$\,meV \cite{Roslawska2022}. We choose the value of $\Omega_{0}$ to ensure that the number of emitted photons roughly corresponds to the experimentally obtained values ($\sim 10^4$ photons/s for excitation at $1.93$\,eV and $2.00$\,eV, and $\sim 10^5$ photons/s at $1.81$\,eV illumination energy). 

In the main text, we consider the values for the tautomerization rates as tip-position-independent and set $\gamma_{\rm AB}^{\rm XA}=\gamma_{\rm AB}^{\rm YA}=\gamma_{\rm AB}^{\rm XB}=\gamma_{\rm AB}^{\rm YB}=\gamma_{\rm AB}^0$. Choosing the tip-position-independent value for $\gamma_{\rm AB}^0$ allows us to simulate the situation in which the tautomerization is mediated by an intermediate state (\textit{e.g.} triplet) that is populated from the singlet excited states. 
In this case, the tautomerization rate originates from the intersystem crossing in the molecule and the tautomerization probability in the triplet-state manifold, both being governed by rates intrinsic to the molecule and independent of the tip position. 
Since we are not able to experimentally distinguish these processes, we include both processes into a single tautomerization rate. 

\begin{figure}[!h]
  \includegraphics[width=\textwidth]{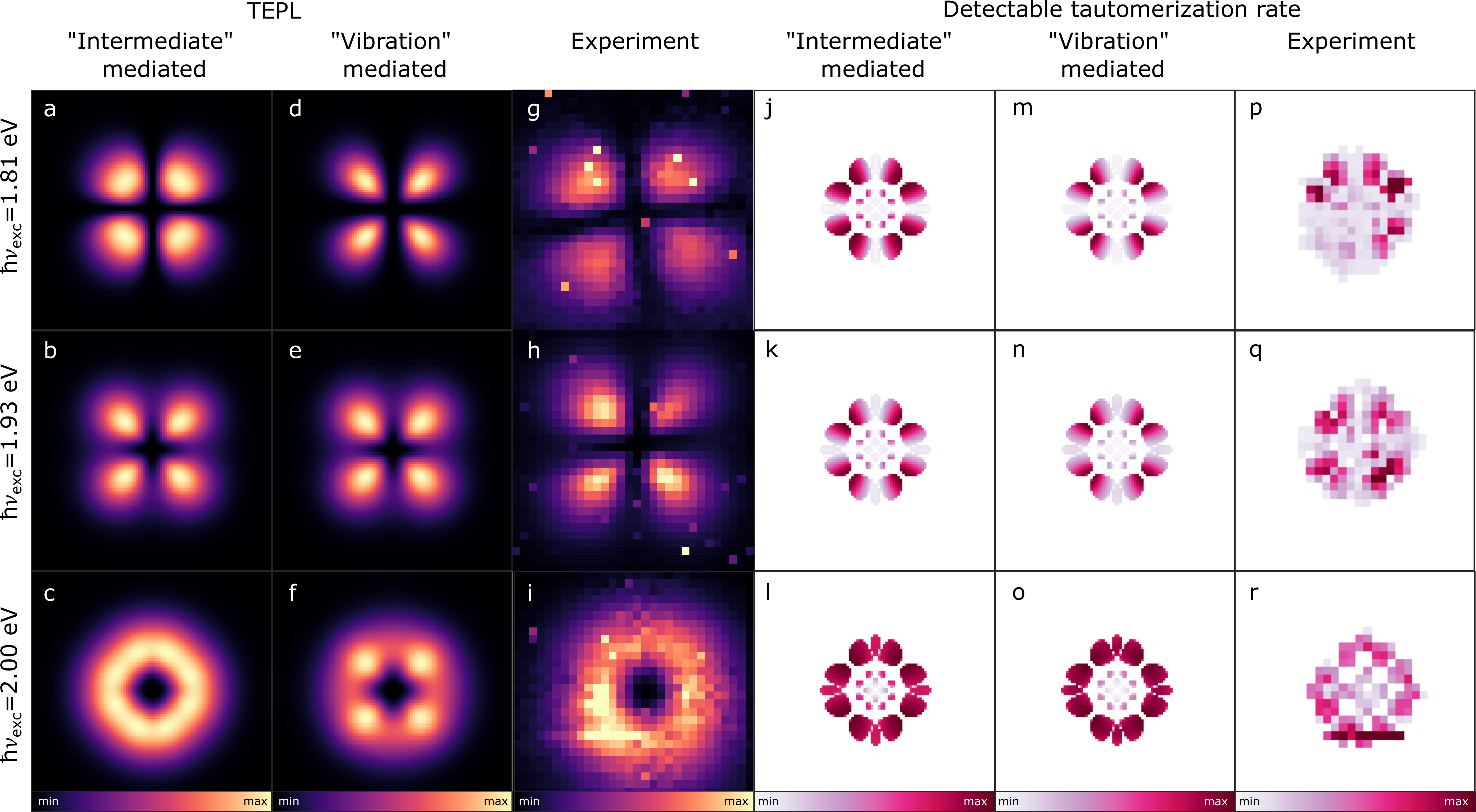}
 \caption{\label{sfig:tripletphonon} TEPL (a-f) and tautomerization (j-o) maps calculated using the model, and the corresponding experimental maps (g-i), and (p-r). In (a-c) and (j-l) we consider the tautomerization rates to be set by a constant value independently of the position of the tip. We label this situation as "intermediate"-mediated tautomerization. On the other hand, in (d-f) and (m-o) we take the tautomerization rates from Eq.\,\ref{seq:tauto1} to Eq.\,\ref{seq:tauto4}. This latter situation is labelled as "vibration"-mediated tautomerization. In (j-r) we set the color of the undetectable tautomerization rate to white.} 
\end{figure} 

To provide further insight into the tautomerization process, we also additionally consider the case when the tautomerization rates are proportional to the decay rates of the singlet excitons and show the result in \Figref{sfig:tripletphonon}. Such a situation would emulate \textit{e.g.} the process where the intermediate state would be the vibrationally excited electronic ground state of the molecule. This state would be populated \textit{e.g.} due to the Franck-Condon activity of the molecular vibrations. In this case, we implement the tautomerization rates as:
\begin{align}
    \gamma_{\rm AB}^{\rm XA}&=\frac{\gamma_{\rm GX}^{\rm A}({\bf r}_{\rm tip})}{\gamma_{\rm GS}+\gamma_{\rm GS}^0}\gamma_{\rm AB}^0,\label{seq:tauto1}\\
    \gamma_{\rm AB}^{\rm YA}&=\frac{\gamma_{\rm GY}^{\rm A}({\bf r}_{\rm tip})}{\gamma_{\rm GS}+\gamma_{\rm GS}^0}\gamma_{\rm AB}^0,\\
    \gamma_{\rm AB}^{\rm XB}&=\frac{\gamma_{\rm GX}^{\rm B}({\bf r}_{\rm tip})}{\gamma_{\rm GS}+\gamma_{\rm GS}^0}\gamma_{\rm AB}^0,\\
    \gamma_{\rm AB}^{\rm YB}&=\frac{\gamma_{\rm GY}^{\rm B}({\bf r}_{\rm tip})}{\gamma_{\rm GS}+\gamma_{\rm GS}^0}\gamma_{\rm AB}^0.\label{seq:tauto4}
\end{align}
When we compare the TEPL maps obtained using the tip-position-dependent rates and the maps obtained with the tip-position-independent rates we conclude that it is more likely that tip-position-independent rates are at play in the experiment.

\begin{center}
\begin{table}
\caption{Parameters used in the theoretical model.}
\begin{tabular}{ c | c | c }\label{tab:param}
 Parameter & Value & Timescale \\ \hline \hline 
 \addstackgap{\shortstack{$\hbar\Omega_0$ \\ Amplitude of optical pumping}} & $0.8$\,$\mu$eV & \addstackgap{$\frac{2\pi}{\Omega_0}\approx$ 13.8 ns} \\  \hline 
 \addstackgap{\shortstack{$\hbar\gamma_{\rm AB}^0$ \\ Intrinsic tautomerization rate}} & $400$\,neV & \addstackgap{$\frac{1}{\gamma_{\rm AB}^0}\approx$ 1.65 ns} \\  \hline
 \addstackgap{\shortstack{$\hbar\gamma_{\rm GS}$ \\ Amplitude of plasmon-induced exciton decay rate}} & 2\,meV  & \addstackgap{$\frac{1}{\gamma_{\rm GS}}\approx$ 0.33 ps} \\  \hline
 \addstackgap{\shortstack{$\hbar\gamma_{\rm GS}^0$ \\ Tip-position-independent exciton decay rate}} & 0.5\,meV & \addstackgap{$\frac{1}{\gamma_{\rm GS}^0}\approx$ 1.32 ps} \\  \hline
 \addstackgap{\shortstack{$\hbar\gamma_{ xy}$ \\ Internal conversion decay rate}} & 1\,meV  & \addstackgap{$\frac{1}{\gamma_{xy}}\approx$ 0.66 ps} \\  \hline
 \addstackgap{\shortstack{$\hbar\gamma_{\rm SS}$ \\ Vibrational decay rate}} & 1\,meV & \addstackgap{$\frac{1}{\gamma_{\rm SS}}\approx$ 0.66 ps}  \\
 \hline
 \addstackgap{\shortstack{$\hbar\gamma_\phi$ \\ Pure-dephasing rate}} & 6\,meV & \addstackgap{$\frac{1}{\gamma_{\phi}}\approx$ 0.11 ps}
\end{tabular}
\end{table}
\end{center}

\newpage

\subsection{S8 -- DFT and TDDFT calculations of transition densities and vibronic spectra}
The transition densities, generalizing the transition dipole moment of the molecule, are calculated using time-dependent density-functional theory (TDDFT) as implemented in Gaussian 16 \cite{g16} using the 6-31G* Gaussian basis set and the hybrid B3LYP functional \cite{becke93}. We calculate the excited states and extract the transition densities in the ground-state equilibrium geometry of the molecule. This geometry is obtained by optimizing the molecular structure in the ground state, as implemented in Gaussian 16. In the model described in Section S7 we do not use the TDDFT energies of Q$_x$ and Q$_y$ excitations and instead use the experimentally observed positions of the respective exciton peaks. 

We next describe the vibronic activity of the molecule. To that end, we first optimize the geometry of the molecule in the ground state, and both electronic excited states corresponding to the Q$_x$ (S$_0$ $\to$ S$_1$) and Q$_y$ (S$_0$ $\to$ S$_2$) transitions in absorption. In each state, we perform the geometry optimization and calculate the vibrational modes. We scale the vibrational energies by a multiplicative factor of 0.95 as the (TD)DFT energies are usually slightly overestimated compared to experimental values. Next, we perform the Franck-Condon analysis by numerically projecting the appropriately quantized mass-weighted vibrational modes into the vector of the mass-weighted difference between the geometry in the ground electronic state and the respective excited state. By doing so we extract the dimensionless displacement of the normal modes in S$_1$ ($D_i^x$), and in S$_2$ ($D_i^y$) with respect to the ground state $S_0$. The model parameter $d^{x(y)}_i$ of mode $i$ is related to $D^{x(y)}_i$ as $d^{x(y)}_i=D^{x(y)}_i/2$. In the calculation we then include modes for which $|d_i^x|>0.01$ and insert them into the model outlined in Section S7. 

We calculate the absorption spectrum by varying the incident laser energy $\omega$ and evaluating the joined population of the molecule's excited states $N_{\rm exc}=\langle\sigma_{\rm XX}\rangle+\langle\sigma_{\rm YY}\rangle$. We plot $N_{\rm exc}$ as a function of $\omega$ in Fig.\,\ref{SI_vib}d.

\newpage

\subsection{S9 -- DFT calculation of orbitals in Octopus}

In the experiment, the tautomerization events are detected as steps in the tunnelling current when bias voltage and tip-sample distance are kept constant. This procedure modifies the spatial distribution of the detected tautomerization rates (tautomerization maps) by imprinting the detection sensitivity into the spatial pattern. Below a certain threshold of current difference, the experiments detect no tautomerization events and thus set the tautomerization rate equal to zero. As a consequence, only the areas of a significant current-difference signal are seen in the experimental images, in contrast to the theoretically calculated images of tautomerization rates that do not consider any particular detection mechanism. We refer the reader to \Figref{SI_histogram}d for the experimental spatial distribution of the difference between tunneling current passing through tautomer A and tautomer B.

To account for this effect, we, therefore, simulate the detection process and calculate the partial density of states $|\psi^{A}_{\rm LUMO}|^2$ related to the lowest unoccupied molecular orbital (LUMO) of H$_2$Pc in the two tautomer configurations $A\in \{{\rm A},{\rm B}\}$ (accounting for the 90$^\circ$ rotation of the orbital). This partial density of states is directly linked to the tunnelling current detected under experimental conditions. We then simulate the sensitivity to the current difference by evaluating $\rho_{\rm diff}=\left ||\psi^{\rm A}_{\rm LUMO}|^2-|\psi^{\rm B}_{\rm LUMO}|^2\right|$ at a constant height ($z=10a_0$, with $a_0$ being the Bohr radius) above the molecular plane ($z=0$). Since the experiment is unable to detect tautomerization below a certain threshold (under which it indicates zero tautomerization rate), we define the region $R({\bf r}_{\rm tip})$ of experimental sensitivity as a function of tip positions ${\bf r}_{\rm tip}$ at a constant height above the molecule such that
\begin{align}
    R({\bf r}_{\rm tip})&=
    \bigg\{
    \begin{array}{c}
    1\;{\rm for}\;\rho_{\rm diff}({\bf r}_{\rm tip})>0.1\times{\rm max}\{ \rho_{\rm diff}({\bf r}_{\rm tip}) \},\\ 
    0\;{\rm for}\;\rho_{\rm diff}({\bf r}_{\rm tip})\le 0.1\times{\rm max}\{ \rho_{\rm diff}({\bf r}_{\rm tip}) \}.
    \end{array}
\end{align}

We evaluate the orbital $\psi_{\rm LUMO}$ by performing a ground-state DFT calculation using Octopus 11.3 \cite{tancogne2020} which implements the self-consistent Kohn-Sham scheme on a real-space grid. We perform the calculation in the Perdew-Zunger \cite{perdew1981} parametrization of the local density approximation (LDA) correlation and the Slater density functional for the LDA exchange functional \cite{dirac1930,slater1951} using the ground-state equilibrium geometry obtained in Gaussian 16. 
We use the "minimum" box shape and set the radius to $10$\,\AA, and spacing to $0.1$\,\AA. We extract the Kohn-Sham orbitals in the form of cube files. 

\newpage

\subsection{S10 -- Tautomerization mechanism}

\begin{figure*}[!h]
  \includegraphics{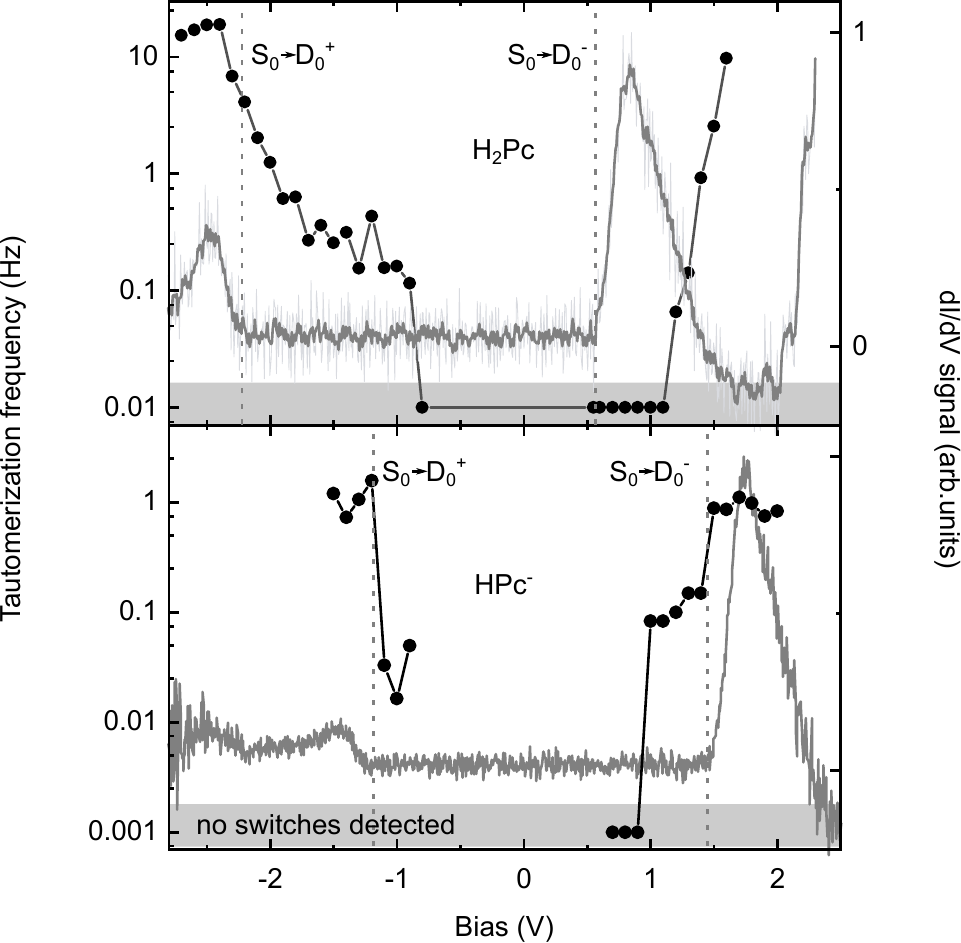}
 \caption{\label{SI_biasdep} Bias dependence of current-induced tautomerization (left axis). The measurement was done in a constant current mode. For the data points in the region indicated by grey rectangles no switches were detected and an artificial offset, necessary to use the logarithmic scale, was added. Upper plot: H$_{2}$Pc, current set-point: 20 pA. Bottom plot: HPc$^{-}$ set-point: 10 pA. The d$I$/d$V$ spectra of both molecules are plotted in grey (right axis).} 
\end{figure*}

As discussed in Section S7, the comparison between the models considering either a reaction path involving the tip-position-dependent exciton decay (\textit{e.g.} tautomerization mediated by ground-state vibrations) or a path involving a tip-position-independent rate suggests that the latter is at play. This reaction path can rely on, for example, an intersystem crossing to the triplet manifold. Here, we discuss experimental data that further support that interpretation.

Tautomerization in both H$_{2}$Pc and HPc$^{-}$ (see Section S2) adsorbed on NaCl can be induced by tunnelling electrons when a high enough bias is applied as presented in \Figref{SI_biasdep} \cite{Doppagne2020, Vasilev2022}. For H$_{2}$Pc, using negative bias, we find three onsets of tautomerization, at bias voltage values close to $V$ = - 1 V, - 1.8 V and - 2.2 V. The latter one corresponds to the transition to the transiently charged D$_0^+$ state (where D$_0^+$ (D$_0^-$) is the ground state of a positively (negatively) charged molecule). Due to its high energy, the molecule can then be neutralized to either the singlet (S$_1$) or triplet (T$_1$) state \cite{Jiang2022}. For electrons of energy between 1.8 and 2.2 eV, the molecule can be inelastically driven to the S$_1$ state. The lowest threshold is close to the onset of the triplet state, estimated to be in the range 1.1-1.2 eV (or even lower) \cite{McVie1978,Frackowiak2001}, which, similarly to S$_1$, can be inelastically excited \cite{Chen2019}. Applying positive bias, since the transition to the D$_0^-$ is at low energy, we find only one threshold for the tautomerization (1.2 eV). Note that the D$_0^+$ or D$_0^-$ energies are deduced from the d$I$/d$V$ spectroscopy (grey trace).

HPc$^{-}$ on a thin NaCl film is charged. However, the charge is located in a $\sigma$-orbital and, besides a rigid shift in energy due to electrostatic interactions, does not influence transport or optical properties that solely depends of the $\pi$-orbitals of the molecule\cite{Vasilev2022}. In the following, we therefore consider that the ground state of the molecule is neutral (S$_0$). For HPc$^{-}$, we observe a sharp increase in tautomerization rate when we can drive the molecule to the D$_0^+$ state. In contrast to H$_{2}$Pc, HPc$^{-}$ cannot be driven to the S$_1$ state from D$_0^+$, and does not electroluminesce when excited directly \cite{Vasilev2022}. Therefore, only the transition to the T$_1$ state is possible, further indicating that the decay of S$_1$ to vibrational states, in principle possible in the case of transition to D$_0^+$ in H$_{2}$Pc, does not drive tautomerization. In analogy to the singlet \cite{Vasilev2022}, we estimate the energy of the triplet state of HPc$^{-}$ to be slightly (50 meV) blueshifted with respect to H$_{2}$Pc. Note that for very low negative bias voltages, we were not able to conduct measurements due to the instability of the molecule on the surface. When applying positive bias we find a threshold close to 1 eV and an increase in the tautomerization signal at the transition to the D$_0^-$ state. Again, as when applying the negative bias, only the transition to the T$_1$ state is possible.

Combined, we find that the energy thresholds of current-induced tautomerization in H$_{2}$Pc and HPc$^{-}$ approximately correspond to the energy of the triplet state. Excitation of the T$_1$ state can be enhanced by including a transiently charged, D$_0^+$ or D$_0^-$, state.

\newpage
\noindent

\bibliographystyle{naturemag}

\bibliography{tauto_ref}